\documentclass{aa}
\usepackage{graphicx}
\usepackage[varg]{txfonts}
\usepackage{multirow}
\usepackage{subcaption}
\usepackage{adjustbox}
\usepackage{amssymb}
\usepackage{xcolor}
\usepackage{colortbl}
\usepackage[colorlinks=true,allcolors=blue]{hyperref}
\newcommand{\grayline}{\arrayrulecolor{violet}\hline\arrayrulecolor{black}}

\graphicspath{{./}{figures/}}

\begin{document}

\title{HYPERION. Interacting companion and outflow in the most luminous $z>6$ quasar}

\titlerunning{Interacting companion and outflow in the most luminous $z>6$ quasar}
\authorrunning{R. Tripodi et al.}

\author{R. Tripodi
    \inst{1,2,3,4,5}\thanks{\email{roberta.tripodi@inaf.it}}
    \and J. Scholtz
    \inst{4,5}
    \and R. Maiolino
    \inst{4,5,6}
    \and S. Fujimoto
    \inst{7,8,9}
     \and S. Carniani
     \inst{10}
    \and J. D. Silverman
    \inst{11,12}
    \and C. Feruglio
    \inst{2,3}
    \and M. Ginolfi
    \inst{13}
    \and L. Zappacosta
    \inst{14}
    \and T. Costa
    \inst{15}
    \and G. C. Jones
    \inst{16}
    \and E. Piconcelli
    \inst{14}
     \and M. Bischetti
    \inst{1,3}
    \and F. Fiore
    \inst{2,3}
    }

 \institute{Dipartimento di Fisica, Università di Trieste, Sezione di Astronomia, Via G.B. Tiepolo 11, I-34131 Trieste, Italy
        \and
        INAF - Osservatorio Astronomico di Trieste, Via G. Tiepolo 11, I-34143 Trieste, Italy
         \and
         IFPU - Institute for Fundamental Physics of the Universe, via Beirut 2, I-34151 Trieste, Italy
         \and
         Institute of Astronomy, University of Cambridge, Madingley Road, Cambridge CB3 0HA, UK
         \and
         Kavli Institute for Cosmology, University of Cambridge, Madingley Road, Cambridge CB3 0HA, UK
         \and 
         Department of Physics and Astronomy, University College London, Gower Street, London WC1E 6BT, UK
         \and
         Department of Astronomy, The University of Texas at Austin, Austin, TX, USA
         \and
         Cosmic Dawn Center (DAWN), Denmark
         \and
         Niels Bohr Institute, University of Copenhagen, Jagtvej 128, DK2200 Copenhagen N, Denmark
         \and
         Scuola Normale Superiore, Piazza dei Cavalieri 7 I-56126 Pisa, Italy
         \and
         Kavli Institute for the Physics and Mathematics of the Universe, The University of Tokyo, Kashiwa, 277-8583, Japan
         \and
         Department of Astronomy, School of Science, The University of Tokyo, 7-3-1 Hongo, Bunkyo, Tokyo 113-0033, Japan
         \and 
         Dipartimento di Fisica e Astronomia, University of Firenze, Via G. Sansone 1, 50019, Sesto Fiorentino (Firenze), Italy
         \and
         INAF - Osservatorio Astronomico di Roma, Via Frascati 33, I-00040 Monte Porzio Catone, Italy
         \and
         Max-Planck-Institut für Astrophysik, Karl-Schwarzschild-Straße 1, D-85748 Garching b. München, Germany
         \and
         Department of Physics, University of Oxford, Denys Wilkinson Building, Keble Road, Oxford OX1 3RH, UK
         \\
             }

\abstract{We present ALMA deep observations of the [CII] 158 $\mu$m emission line and the continuum at 253 GHz and 99 GHz towards SDSS J0100+2802 at $z\simeq 6.3$, the most luminous quasi-stellar object (QSO) at z$>$6. J0100+2802 belongs to the HYPERION sample of luminous QSOs at $z\sim 6-7.5$. The observations have a 2.2 arcsec resolution in band 3 and a 0.9 arcsec resolution in band 6, and are optimized to detect extended emission around the QSO. We detect an interacting, tidally disrupted companion both in [CII], peaking at $z\sim 6.332$, and in continuum, stretching on scales up to 20 kpc from the quasar, with a knotty morphology.  The higher velocity dispersion in the direction of the companion emission and the complex morphology of the tidally stretched galaxy suggest a possible ongoing or future merger. For the newly detected companion, we derive the range of the dust mass, $M_{\rm dust}=(0.3-2.6)\times 10^7\ \rm M_\odot$, and of the star formation rate, SFR$=[35-344]\ \rm M_\odot$, obtained from the modelling of its cold dust spectral energy distribution. This shows that both the QSO and its companion are gas-rich and that a major merger may be at the origin of the boosted star formation. This close interacting companion is undetected by deep JWST imaging observations, showing the effectiveness of ALMA in detecting dust-obscured sources, especially in the vicinity of optically bright quasars. We also detect a broad blueshifted component in the [CII] spectrum, which we interpret as a gaseous outflow for which 
 we estimate a mass outflow rate in the range $\dot{M}_{\rm out}=(118-269)\ \rm M_\odot\ yr^{-1}$. 

J0100+2802 was recently found to reside in a strong overdensity, however this close companion remained undetected by both previous higher resolution ALMA observations and by JWST-NIRCAM imaging.
Our results highlight the importance of deep medium-resolution ALMA observations for the study of QSOs and their environment in the Epoch of Reionisation. 
}

   
\maketitle

\begin{table*}[t]
\caption{Summary of the ALMA observations and their properties }    
\centering       
\label{table:obs}  
\resizebox{2\columnwidth}{!}{%
\begin{tabular}{c c c c c c c c}        
\hline\hline  
Band & Project ID & RA, DEC & Central Freq. & L5 BL, L80 BL & Synth. beam & R.m.s. cont. & R.m.s. cube  \\
 & & (J2000) & (GHz) & (m) (m) & (arcsec$^2$) & ($\mu$Jy/beam) & (mJy/beam) \\ 
\hline
 3 & 2021.1.00211 &  01:00:13.024, +28:02:25.790 & 99.48 & 26, 184 & 2.59$\times$1.95  &   5 & 0.1  \\
6 & 2021.1.00211 &  01:00:13.024, +28:02:25.790 & 252.87 & 26, 182 &  1.08$\times$0.82 & 10 & 0.2 \\ 
\hline
                                 
\end{tabular}}
 \flushleft 
\footnotesize {{\bf Notes.} Columns are: (1) ALMA band; (2) project ID of the observation; (3) coordinates of the pointing of the observation; (4) central frequency of the spectral setup; (5) lengths that include the 5th and the 80th percentile of all projected baselines calculated from the unweighted UV distribution; (6) synthesized beam of the observation for the continuum data; (7) r.m.s of the continuum map; (8) r.m.s of the data cube that includes the CO(6-5) and [CII] emission line for band 3 and band 6, respectively.}
\end{table*}

\section{Introduction}

The presence of luminous quasi-stellar objects (QSOs) near the end of the Epoch of Reionisation (EoR), when the Universe was only 0.5-1 Gyr old, is currently a puzzle and represents a real challenge to theoretical models aiming to explain supermassive black hole (SMBH) formation and growth on such a short timescale \citep[e.g.,][]{volonteri2010, johnson2016, greene2020, kroupa2020, inayoshi2020, lupi2021, fan2022, trinca2022, volonteri2023}. Moreover, the well-established correlations between the black hole (BH) mass and the physical properties of the host galaxy \citep{kormendy2013} strongly suggest that the growth of these SMBHs has to be connected with the growth of their host galaxies, and thus with the properties of the interstellar medium (ISM) \citep[e.g.,][]{dimatteo2005, hopkins2008, harrison2018}. In this picture, extreme physical processes, such as mergers and active galactic nuclei (AGN) feedback, play a critical role in shaping the evolution and the properties of the host galaxies, especially at high redshift when these objects are caught in the first phases of their formation. In the last decade, massive AGN-driven outflows have been observed, involving different gas phases (ionised, atomic, and molecular) extending from sub-parsec to kiloparsec scales up to high redshift ($z\sim 1-7$, e.g. \citealt{fiore2017, carniani2017, feruglio2017, brusa2018, bischetti2019,vietri2022}). 

The study of the [CII] fine-structure emission line at $158\ \mu$m, especially with the Atacama Large Millimetre Array (ALMA), has brought significant advances in the understanding of the host galaxies' properties. This is one of the brightest emission lines in QSO host galaxies at far infrared (FIR) wavelengths and arises from photodissociation regions (PDRs; \citealt{hollenbach1999}) at the interface of the atomic and molecular media on the outskirts of molecular clouds in galactic star-forming regions, and therefore is an optimal diagnostic for studying the ISM in the cold, warm neutral, and mildly ionised phases \citep{cormier2015,olsen2018}. \citet{stanley2019} and \citet{bischetti2019} both find very broad [CII] wings that trace the presence of outflows, with a velocity excess of up to $1000$ km s$^{-1}$, performing a stacking analysis of a sample of 20 QSOs at $z\sim 6$ and 48 QSOs at $4.5<z<7.1$, respectively. \citet{novak2020} find no evidence for broad wings with a similar stacking technique; however, in contrast to \citet{bischetti2019}, they used high angular resolution observation, which may have filtered out the extended outflow component.  Very broad wings were also observed in the hyper-luminous QSO SDSS J1148+5251 at $z\sim 6.4$ \citep{maiolino2012, cicone2015}, revealing outflowing gas extended up to $\sim 30$ kpc, although that result has been disputed by \citet{meyer2022}.
Other indications of outflows in z$>$6 quasars through [CII] broad wings have been found by \citet{izumi2021a,izumi2021b}.
Moreover, the powerful capabilities of ALMA have allowed a detailed analysis of the kinematics and dynamics of the gas through the observation of the [CII] emission line, revealing evidence of rotating discs \citep[e.g.,][]{pensabene2020, neeleman2021, tsukui2021, shao2022, roman2023} and bulges \citep{lelli2021, tripodi2023a} in $z>4$ QSO host galaxies. The characterisation of the dynamical state of galaxies at high-z is indeed crucial for determining the mechanisms of mass assembly in the early Universe. In this regard, the study of QSO samples specifically designed to tackle the properties of luminous $z>6$ quasars, powered by the SMBHs that experienced the fastest mass accretion growth, may offer a physical framework to interpret the dynamical states of the host galaxies. The HYPerluminous quasars at the Epoch of ReionizatION survey (HYPERION; \citealt{zappacosta2023}) includes the titans among $z>6$ QSOs, that is, those in which the SMBHs acquired the largest masses at their epoch, possibly resulting from extreme concurrent nuclear or host galaxy accretion and dynamical interaction pathways.

In this work we present ALMA band 3 and band 6 observations of the HYPERION QSO SDSS J010013.02+280225.8 (hereafter J0100+2802) at $z_{\rm [CII]}=6.327$ \citep{wang2019}. \citet{wu2015} estimated a bolometric luminosity of $L_{\rm bol}=4.29\times 10^{14}\rm \ L_{\odot}$ and a BH mass of $M_{\rm BH}=1.24\times10^{10}\ \rm M_{\odot}$ for J0100+28, making it the most optically luminous QSO with the most massive SMBH known, at $z>6$. James Webb Space Telescope (JWST) observations by \citet{eilers2022} have confirmed that this QSO is not lensed. This QSO is also found to reside in a strong overdensity, traced by 24 galaxies \citep{kashino2023}. One or more galaxies were identified within 200 pkpc and 105 km s$^{-1}$ of four metal-absorption systems, either as their physical host galaxies or as neighbouring gas-galaxy associations embedded in a common larger-scale structure.

The properties of the cold ISM of J0100+2802 have been studied by \citet{wang2019}, who measured the CO spectral line energy distribution. They find the molecular gas has a mass, $M_{\rm H_2}=(5.4 \pm 1.6) \times 10^9\ \rm M_\odot$, characterised by two primary phases, a cool one at $\sim 24^{+8}_{-3}$ K with $n_{\rm H_2}=10^{4.5\pm 1.1}$ cm$^{-3}$, and a warm one at $\sim 224^{+165}_{-100}$ K with $n_{\rm H_2}=10^{3.6^{+1.3}_{-0.8}}$ cm$^{-3}$. \citet{tripodi2023b} measured the star formation rate (SFR) of the host galaxy with a high accuracy, SFR$=265 \pm 32\ \rm M_\odot$ yr$^{-1}$, studying the cold dust spectral energy distribution (SED). \citet{neeleman2021}, through the analysis of the [CII] emission line, did not find any sign of a velocity gradient and concluded that the system was possibly dominated by turbulent motions. Finally, \citet{fujimoto2020} and \citet{novak2020} report a complex, multi-clump morphology in the dust continuum at a scale of 0.5 arcsec$\sim 3$ kpc. The origin is unclear, but it suggests the possibility of dusty mergers or companions or a gravitationally lensed system, where the latter has been ruled out \citep{eilers2022}.

In this paper we present very high-sensitivity ALMA data of J0100+2802, revealing for the first time in this source clear velocity gradients indicative of the presence of an interactive companion and a [CII] outflow.

\begin{figure*}
    \centering
    \includegraphics[width=0.85\linewidth]{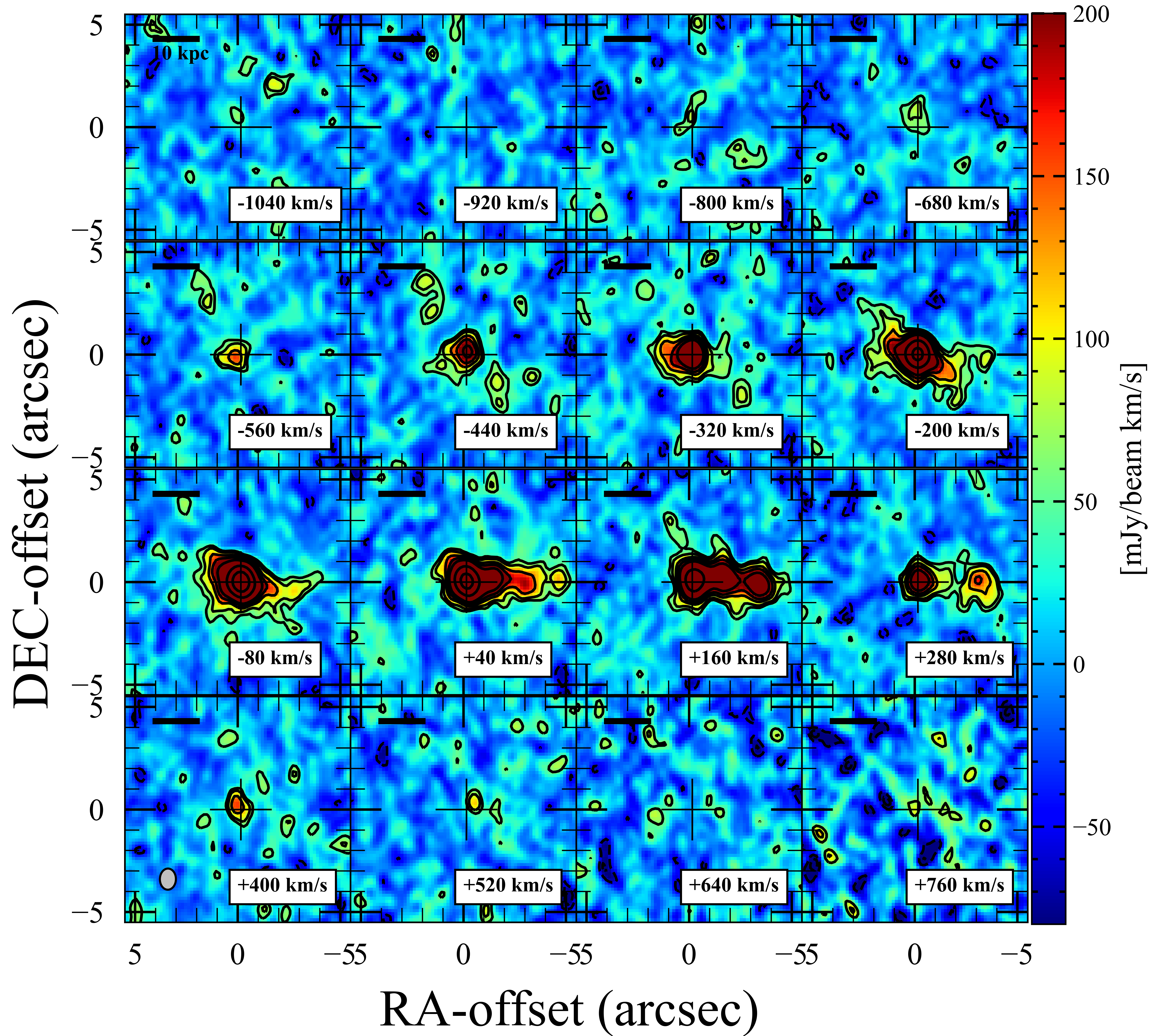}
    \caption{Channel maps of the [CII] emission line for J0100+2802. The cross indicates the peak position of the continuum and velocities are relative to the redshift of the [CII], as determined in \citet{wang2019}. Contours are at $-3,-2,2,3,5,7,10,25,\ {\rm and}\ 50\sigma$. The clean beam is shown in the inset in the lower left corner.}    
    \label{fig:3}
\end{figure*} 

\begin{figure}
    \centering
    \includegraphics[width=1\linewidth]{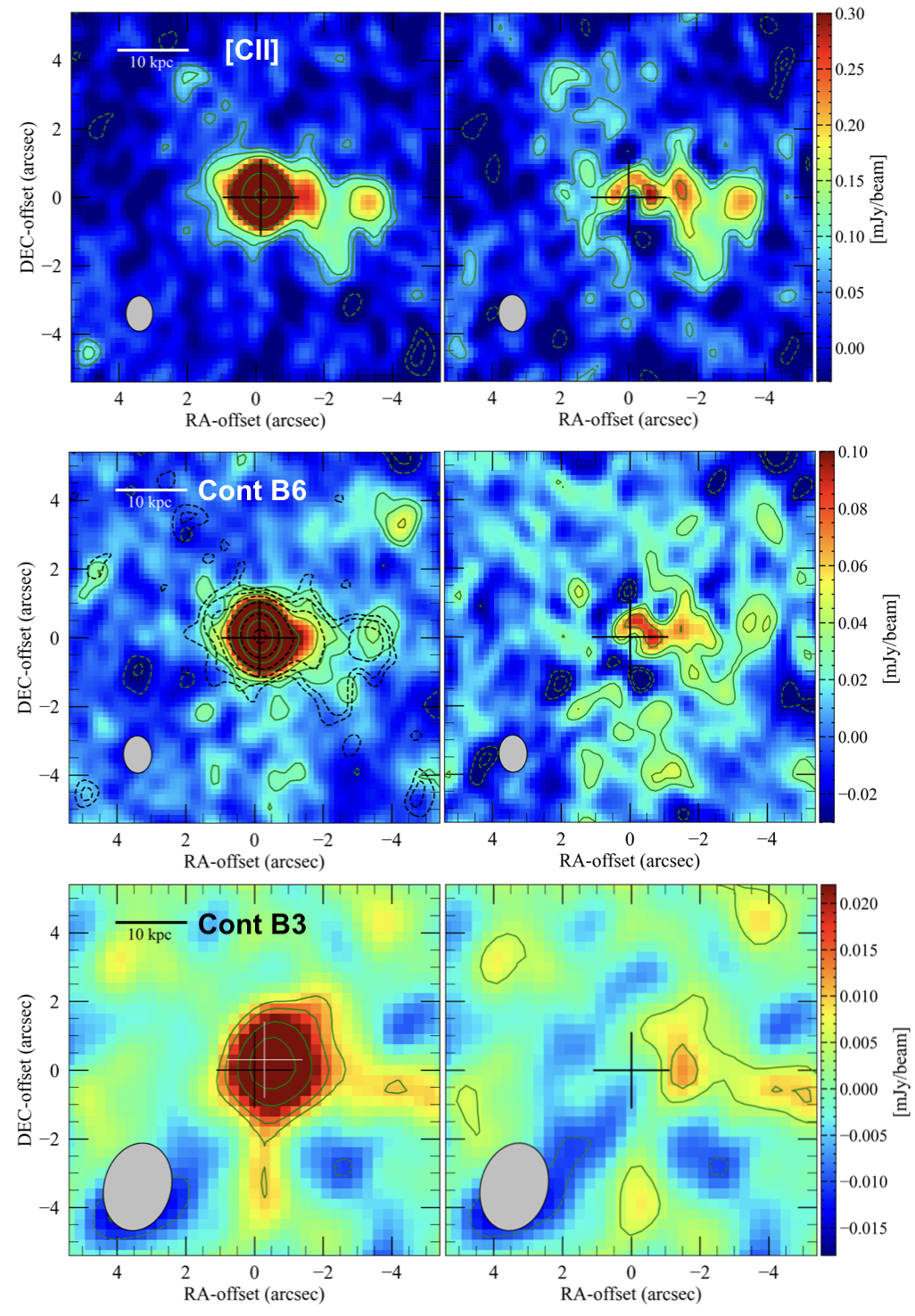}
    \caption{[CII] and continuum maps of J0100+2802. In each panel, the clean beam is plotted in the lower left corner and the cross indicates the peak position of the QSO continuum in band 6. Top panels: [CII] map with green contours at $-3,-2,2,3,5,10,25,\ {\rm and}\ 50\sigma$, with $\sigma=0.03$ mJy/beam (left). PSF-subtracted [CII] map with green contours at $-3,-2,2,3,\ {\rm and}\ 5\sigma$, with $\sigma=0.03$ mJy/beam (right). Central panels: continuum map in band 6 with green contours at $-3,-2,2,3,5,10,25,\ {\rm and}\ 50\sigma$, with $\sigma=0.01$ mJy/beam (left). [CII] contours are overplotted as a dashed black line at $-3,-2,2,3,5,10,25,\ {\rm and}\ 50\sigma$, with $\sigma=0.03$ mJy/beam. PSF-subtracted continuum map with green contours at $-3,-2,2,3,\ {\rm and}\ 5\sigma$, with $\sigma=0.01$ mJy/beam (right). Bottom panels: continuum map in band 3 with green contours at $-3,-2,2,3,5,\ {\rm and}\ 7\sigma$, with $\sigma=0.005$ mJy/beam (left). PSF-subtracted continuum map with green contours at $-3,-2,2,\ {\rm and}\ 3\sigma$, with $\sigma=0.005$ mJy/beam (right). The white cross marks the peak of the continuum in band 3.}
    \label{fig:1}
\end{figure}

\begin{figure}
    \centering
    \includegraphics[width=1\linewidth]{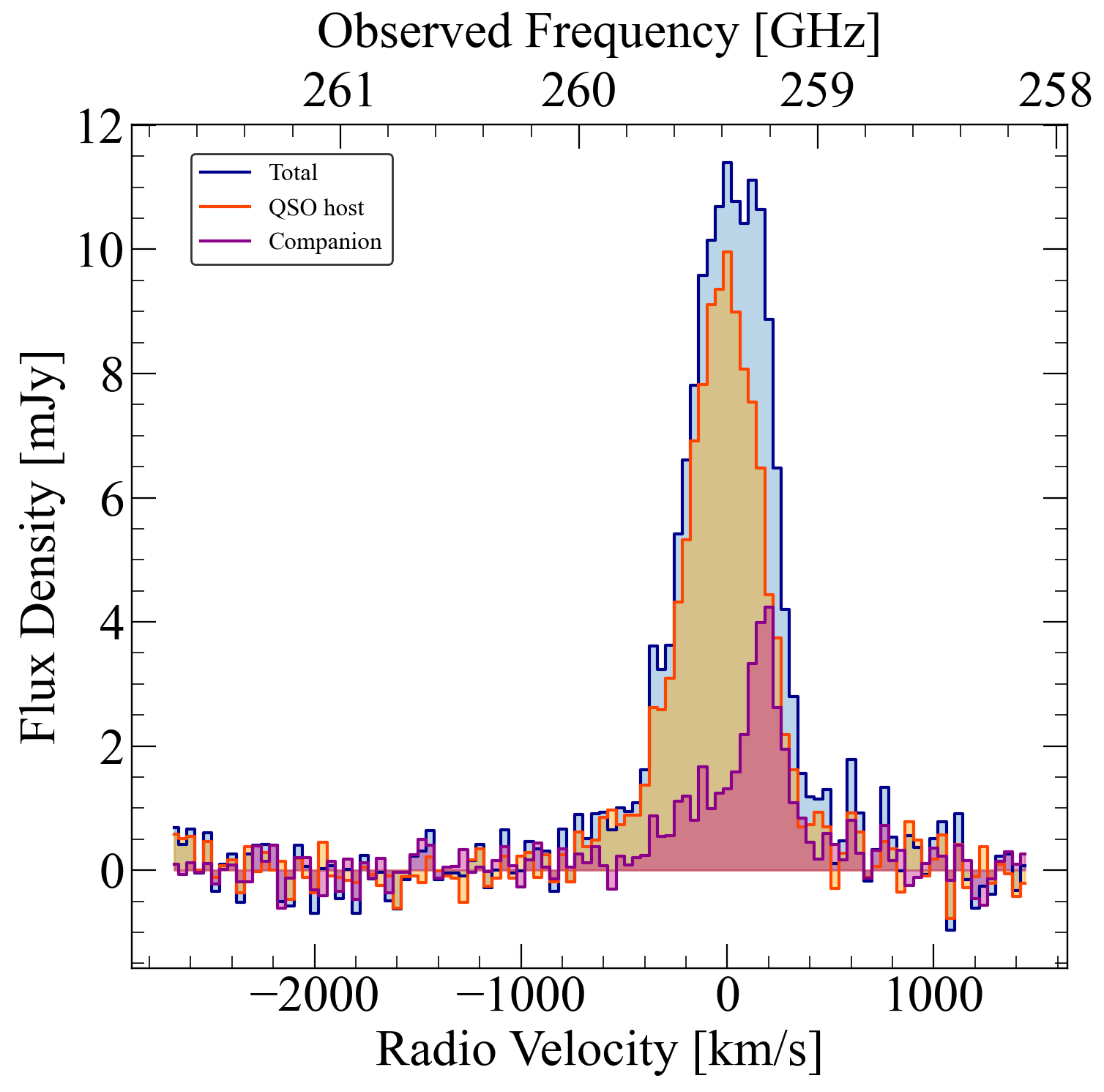}
    \caption{Spectra extracted from three different spatial regions: from a region with a S/N$>2$ in blue; from a region with a S/N$>2$ and RA-offset $<-1.3$ arcsec in purple; and from a region with a S/N$>2$ and RA-offset $>-1.3$ arcsec in orange. We set the 0-velocity reference of the spectrum as for the spectrum in Fig. \ref{fig:4} (see Sect. \ref{sec:outflow}).}
    \label{fig:spec}
\end{figure}

The paper is organised as follows: in Sect. \ref{sec:obs} we describe the observation; in Sect. \ref{sec:analysis} we report the analysis and results; in Sect. \ref{sec:disc} we discuss the implications of our results; and in Sect. \ref{sec:summ} we summarize our results. Throughout the paper, we adopt a $\Lambda$CDM cosmology from \citet{planck2018}: $H_0=67.4\ \rm km\ s^{-1}\ Mpc^{-1}$, $\Omega_m = 0.315$, and $\Omega_{\Lambda} = 0.685$. Thus, the angular scale is $5.66$ kpc/arcsec at $z=6.3$.

\section{Observation}
\label{sec:obs}

We analysed the dataset 2021.1.00211.S (PI: R. Maiolino) from the ALMA 12m array, designed to detect the [CII] and CO(6-5) emission lines of J0100+2802 and their underlying continuum emission at 252.8 GHz (band 6) and 99.5 GHz (band 3). Observations were carried on 21 April 2022 and 21 January 2021, with an integration time of 3.5 hours and 7.5 hours for band 6 and band 3, respectively. The visibility calibration of the observations was performed by the ALMA science archive. 
The imaging was performed through the Common Astronomy Software Applications (CASA; \citealt{casa2022}), version 5.1.1-5. To image the source, we applied \texttt{tclean} using natural weighting and a $3\sigma$ cleaning threshold. We imaged the continuum in band 6 using the multi-frequency synthesis (MFS) mode in all line-free\footnote{[CII] is detected in channels 90-240 for spw2 and 0-40 for spw3.} channels, selected by inspecting the visibilities in all four spectral windows. The r.m.s noise reached for the continuum is $0.01$ mJy/beam, and we obtained a clean beam of ($1.03\times 0.78$) arcsec$^2$, and a position angle, PA = 0.96$^\circ$. Since the [CII] show emission up to very high velocities ($\sim 1000$ km s$^{-1}$, see Sect. \ref{sec:outflow}), we combined the two spectral windows in the upper band to ensure a reliable continuum subtraction, and [CII] is now detected in channels 90-265, corresponding to 1584 km s$^{-1}$. We used the CASA task \texttt{uvcontsub} to fit the continuum visibilities in the line-free channels with a zeroth-order polynomial. We obtained a continuum-subtracted cube with spectral channels of width 9 $\rm km\ s^{-1}$, a clean beam of ($1.08\times 0.82$) arcsec$^2$, PA=1.7$^\circ$, corresponding to an effective radius of $\sim 0.57$ arcsec $\simeq 3.23$ kpc, and an r.m.s noise of $0.2$ mJy/beam per channel. 

We imaged the continuum in band 3 using the MFS mode in all line-free\footnote{CO(6-5) is detected in channels 95-145 for spw3} channels, selected by inspecting the visibilities in all four spectral windows. The r.m.s noise reached for the continuum is $5\ \mu$Jy/beam, and we obtained a clean beam of ($2.59\times 1.95$) arcsec$^2$, PA=15.2$^\circ$, corresponding to an effective radius of $\sim 1.35$ arcsec $\simeq 7.6$ kpc. The analysis of the CO(6-5) did not deliver any improvements with respect to the results published in \citet{wang2019}, therefore here we report only the results on its underlying continuum emission.

The high sensitivity of these observations enabled us to perform a detailed analysis of the [CII] emission and of the continua unveiling a cold outflow and an interactive companion. Moreover, we were able to spatially resolve the outflowing emission and to separate if from the total [CII] emission line profile.

The properties of the ALMA observations used in this work are summarized in Tab. \ref{table:obs}.

\begin{figure*}
    \centering
    \includegraphics[width=0.8\linewidth]{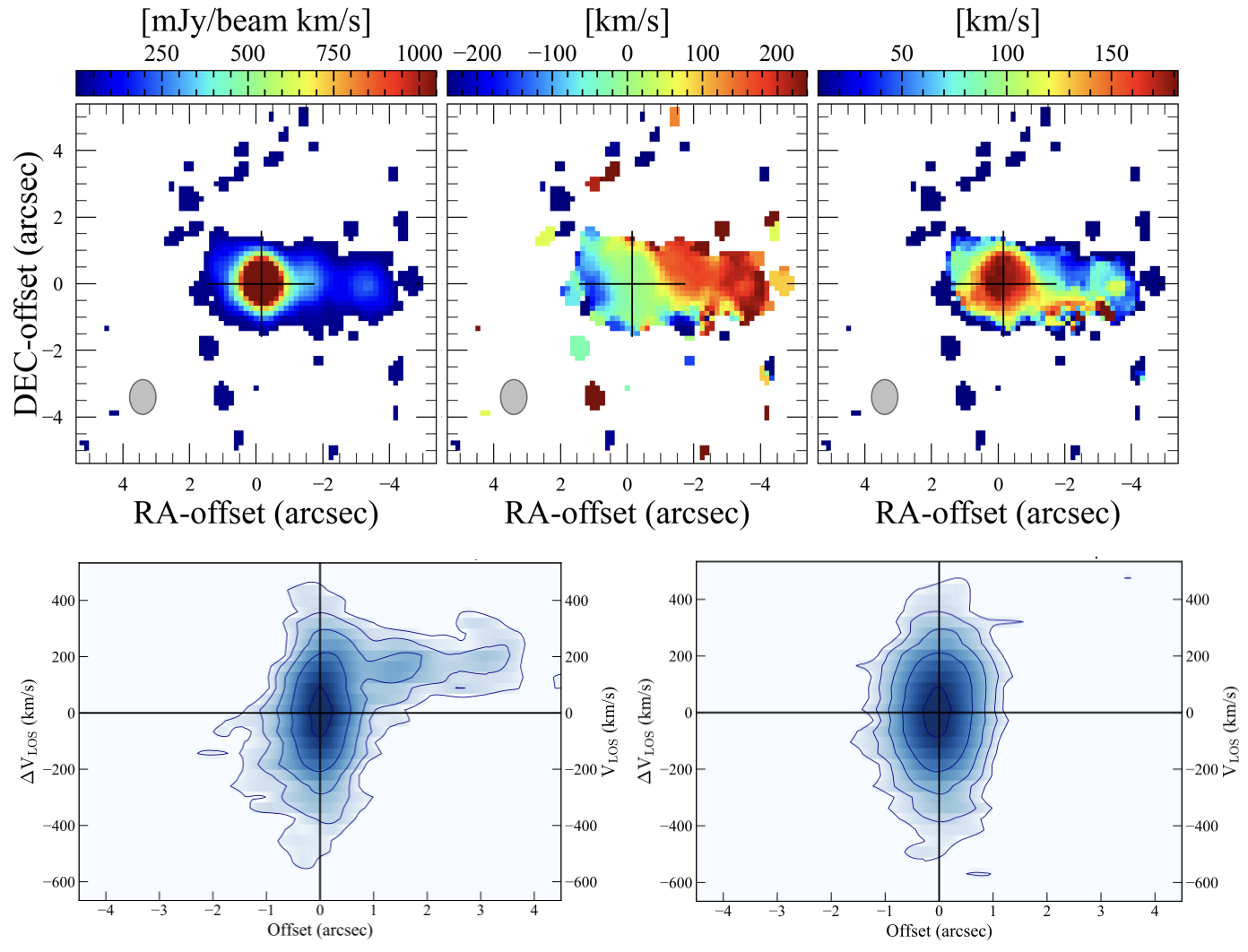}
    \caption{Moment maps and PV diagram of the [CII] emission line. Top panel: integrated flux, mean velocity map, velocity dispersion map, and continuum-subtracted spectrum of [CII]. The clean beam is plotted in the lower left corner of the moment maps. The cross indicates the peak position of the integrated flux. Bottom panel: PV diagrams of the [CII] emission line along the line of nodes (PA=$275^{\circ}$) and orthogonal to it (PA=$185^{\circ}$).} 
    \label{fig:2}
\end{figure*}

\section{Analysis and Results}
\label{sec:analysis}

\subsection{[CII] emission and kinematics}
\label{sec:[CII]}

We used the continuum-subtracted data cube to study the [CII] line emission of the QSO with natural weighting. Details of the complex kinematics of this system can be seen in the channel maps of Fig. \ref{fig:3}, obtained collapsing bins of $120$ km s$^{-1}$ from the continuum-subtracted cube. A prominent, elongated feature is clearly seen at velocities between +40 and +280 km s$^{-1}$. Although it could be interpreted as a jet or an outflow, the fact that it is also detected in the continuum, as discussed in the next section, indicates that it is likely tracing a star-forming companion, extending on scales of $\sim$4 arcsec, in other words  $\sim$20~kpc, that is interacting with the QSO host galaxy. As inferred by the channel maps, the companion is highly structured, with clumps at different velocities. There is also a nearly symmetric and smaller extension on the opposite side and negative velocities (-320 km s$^{-1}$), which is likely tracing a tidal tail of the same interacting system.

The top left panel of Fig. \ref{fig:1} shows the [CII] map, imaged with the MFS mode in the velocity range from $\sim -1000$ to $+600$ km s$^{-1}$, that presents the elongated and complex structure already seen in the channel maps, tracing the interacting companion. Performing a 2D Gaussian fit in a region enclosing the QSO host emission (S/N$>3$, RA:[-2,1.5] arcsec, DEC:[-1.3, 1.5] arcsec), we find a peak flux of $1.52\pm 0.06$ mJy/beam and a flux density of $2.78\pm 0.17$ mJy. 

Even though the fully resolved and tidally stretched structure of the companion is clearly revealed in the channel maps, especially at $v=+40$ km s$^{-1}$ and $v=+280$ km s$^{-1}$, in the [CII] flux map and in the PV diagram, we performed a PSF subtraction to better quantitatively assess the flux of the companion, especially for the continuum emission (see Sect. \ref{sec:cont}). We first computed the flux density from the [CII] map within a region with S/N$>2$ (see $2\sigma$ contours in Fig. \ref{fig:1}), obtaining $3.8\pm 0.05$ mJy. Then, we subtracted from the [CII] map the PSF normalized to the QSO host peak flux. 
The flux density of the companion was computed from the PSF-subtracted [CII] map (top right panel of Fig. \ref{fig:1}) in a region with S/N$>2$ and RA-offset<-1.3 arcsec, to exclude the ring-like feature in the centre that can still be associated with the QSO host, and it is $1.00\pm 0.04$ mJy. Considering this averaged flux within the [CII] line width, we derived $L_{\rm [CII], comp}=(1.68\pm 0.06)\times 10^9\ \rm L_\odot$ (using Eq. 1 in \citealt{solomon2005}). Consequently, the estimated flux density associated with the QSO emission is $2.8\pm 0.07$ mJy, which is consistent with the flux estimated from the 2D fit and which corresponds to a luminosity of $L_{\rm [CII], host}=(4.70\pm 0.12)\times 10^9\ \rm L_\odot$\footnote{$L_{\rm [CII], comp}$ and $L_{\rm [CII], host}$ are computed considering that the [CII] line width is $\sim$1600 km s$^{-1}$.}. This is $\sim 1.25$ times higher than the luminosity derived in \citet{venemans2020}, and this discrepancy is likely due to the higher sensitivity and lower resolution of our observation, which enabled us to observe the fainter and more diffuse [CII] emission, and the high-velocity emission.
The [CII] emission of the QSO is not resolved, while, if considering the whole system, the full width at half maximum (FWHM) size is ($0.89\pm 0.09$)$\times (0.70\pm 0.12)$ arcsec$^2$, corresponding to $\sim (5.03 \times 3.96)$ kpc$^2$ at the rest frame. The PSF-subtracted [CII] map presents multiple peaks towards the west and they arise from the clumps at different velocities that are clearly seen in the channel maps. We may interpret these as the complex morphology of a tidally stretched galaxy in the merging process. This makes the position of the companion difficult to determine precisely and, in principle, these features can also arise from the presence of multiple companions or gas clumps. 
However, the resolution of our observation prevents us from distinguishing between these different scenarios. To be conservative, we associated all the elongated features at S/N$>2$ with a single interactive companion. Spectra extracted from three different spatial regions are presented in Fig. \ref{fig:spec}: the spectrum in blue is extracted from a region with S/N$>2$ including both the QSO and the companion contribution; the one in purple is extracted from a region with S/N$>2$ and RA-offset $<-1.3$ arcsec to better isolate the contribution of the companion to the [CII] emission; and the one in orange is extracted from a region with S/N$>2$ and RA-offset$>-1.3$ arcsec that mostly encloses the QSO host contribution. The [CII] emission arising from the companion peaks at 259.220 GHz (i.e., corresponding to $z=6.3317$), slightly redshifted with respect to the QSO host emission that peaks at 259.378 GHz. 

The top panels of Fig. \ref{fig:2} show the moment-0, -1, and -2 maps of the [CII] emission obtained by applying a $3\sigma$ threshold to the continuum-subtracted cube. The moment-1 map shows a gradient towards the direction of the companion. The moment-2 map shows a range of the velocity dispersion between 10 and 180 $\rm km\ s^{-1}$, where the maximum value towards the nucleus is affected by beam smearing \citep{davies2011}. The left and right bottom panels of Fig. \ref{fig:2} present the position-velocity (PV) diagrams of the disc along the major and minor kinematic axes, respectively. The PA of the major axis is 275$^\circ$. The PV diagram shows an asymmetric structure that partially mimics the S-shape commonly seen in rotating discs. We discuss possible interpretations of these results in Sect. \ref{sec:disc}.

\subsection{Continuum emission of the interacting system}
\label{sec:cont}
The middle left panel of Fig. \ref{fig:1} shows the 253 GHz dust continuum map. The continuum shows an elongated structure westwards, coincident with the [CII] elongation (black contours, see Sect. \ref{sec:[CII]}). Analogously to the [CII] analysis, we performed a 2D Gaussian fit in the region enclosing the QSO emission, and we derived a peak flux of $0.81\pm 0.02$ mJy/beam and a flux density of $1.16\pm 0.05$ mJy, which is consistent with the flux of $1.26\pm 0.08$ mJy found by \citet{wang2019}. We computed the flux density from the continuum map in a region with S/N$>2$, obtaining $1.44 \pm 0.05$ mJy, and we subtracted the point spread function (PSF) normalized to the QSO peak flux from the continuum map. The emission in the PSF-subtracted continuum extends for $\sim$4 arcsec from the centre westwards (see central right panel of Fig. \ref{fig:1}), a similar extension and clumpiness as the [CII] emission. This supports the scenario in which the extended [CII] emission is associated with an interacting system and not an outflowing or jetted component. The flux density obtained in the PSF-subtracted continuum map inside a region with S/N$>2$ and RA-offset $<-1.3$ arcsec is $0.25\pm 0.01$ mJy. Therefore, the QSO continuum flux is $1.19\pm 0.06$ mJy, consistent with the one computed from the 2D fit, and is unresolved at our resolution. The whole emission from the interacting system is spatially resolved with a FWHM size of ($0.68\pm 0.09$)$\times (0.54\pm 0.12)$ arcsec$^2$, corresponding to $\sim (3.84 \times 3.05)$ kpc$^2$ at the rest frame. 

We performed a similar analysis for the continuum emission at 99.5 GHz (bottom left panel of Fig. \ref{fig:1}) in order to determine whether the companion emission was detected. By doing a 2D Gaussian fit, we find that the source has a peak flux of $0.04\pm 0.004$ mJy/beam and a flux density of $0.05\pm 0.007$ mJy. The morphology of the emission is asymmetric, slightly elongated westwards, and coincident with the interacting companion. The elongated feature is less evident than in the band 6 continuum, since the resolution in band 3 is $\sim 2$ times lower. We subtracted the PSF normalized to the QSO peak flux, and indeed we found emission at $3\sigma$ in the same region of the companion emission in band 6. In this case, the emission seems to extend up to -5 arcsec RA-offset from the centre, however this could be an artefact caused by the low resolution. Therefore, we conservatively extracted the continuum flux of the companion from the $2\sigma$ contours enclosed in the region with RA-offset$>-4$ arcsec, obtaining $0.007\pm 0.002$ mJy. 

\begin{table}
\vspace{0.2cm}
		\caption{Properties of QSO host and companion}
		\centering
		\begin{tabular}{llcc}
			\hline
             & & QSO & Companion \\
			\hline
            $S_{\rm cont,\ 253 GHz}$ & [mJy] & $1.19 \pm 0.06$ & $0.25 \pm 0.01$ \\
            $S_{\rm cont,\ 99 GHz}$ & [mJy] & $0.043 \pm 0.005$ & $0.007 \pm 0.002$ \\ 
            $L_{\rm [CII]}^{(*)}$ & [$10^9\ \rm L_\odot$] & 4.70 & $1.68$\\[0.05cm]
            $M_{\rm HI}$ & [$10^9\ \rm M_\odot$] & 6.43 & 2.30\\[0.05cm]
			$T_{\rm dust}$ & [K] & 48$\pm 2$ & 30-50\\[0.05cm]
            $M_{\rm dust}$ & [$10^7 \rm M_\odot$] & 2.3$\pm 0.8$ & 0.3-2.6 \\[0.05cm]
            $\beta$ & & 2.6$\pm 0.2$ & 2.0-3.1 \\[0.05cm]
            SFR & [$\rm M_\odot\ yr^{-1}$] & 265$\pm 32^{(**)}$ & 35-344\\[0.05cm]
			
			\hline
		\end{tabular}
		\label{tab:qso-comp}
		\flushleft 
		\footnotesize {{\bf Notes.} The dust properties and SFR of the QSO host galaxy are taken from \citet{tripodi2023b}. $^{(*)}$: $L_{\rm [CII]}$ were computed inside a region $>2\sigma$ from the [CII] maps shown in the top panels of Fig. \ref{fig:1}, as explained in Sect. \ref{sec:[CII]}. $^{(**)}$ The SFR of the QSO host galaxy was corrected by a factor of 50\% to account for the possible contribution of the AGN to the dust heating \citep{tripodi2023b}.}
	\end{table}
 
\subsection{Dust properties of the companion}

The dust SED of the QSO has previously been studied with high accuracy by \citet{tripodi2023b}, who find a dust temperature of $48 \pm 2$ K, a dust mass of $(2.3 \pm 0.8) \times 10^7\ \rm M_{\odot}$, and an SFR$=265\pm 32 \ \rm M_{\odot}\ yr^{-1}$\footnote{Note that the SFR derived in \citet{tripodi2023b} is corrected by a factor of 50\%, taking into account the contribution of the AGN to the dust heating.}. 
In both low and high frequency previous observations there were no signatures of a companion, probably due to a combination of low sensitivity and high resolution, which were filtering out the extended emission \citep{wang2019, tripodi2023b}. 

\begin{figure}
    \centering
    \includegraphics[width=0.95\linewidth]{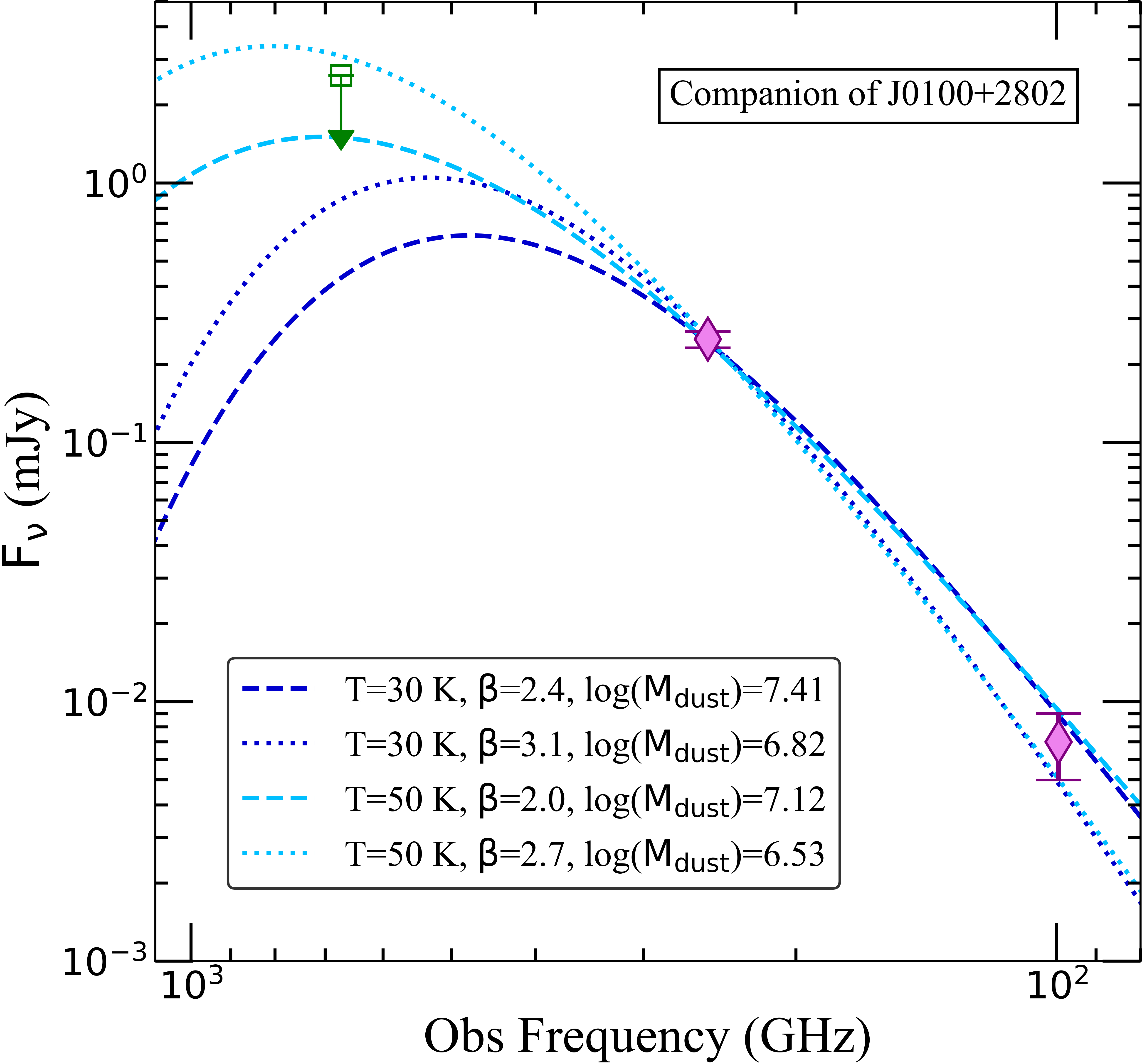}
    \caption{SED of the interacting companion of J0100+2802, using our ALMA data at 252.8 GHz and 99.5 GHz. The best-fitting curve at fixed $T_{\rm dust}=50$ K and $\beta=2.0$ ($\beta=2.7$) is shown as a dashed (dotted) cyan line; the curve at fixed $T_{\rm dust}=30$ K and $\beta=2.4$ ($\beta=3.1$) is shown as a dashed (dotted) blue line. The hollow green square is the upper limit derived in band 9. The estimated value of the dust mass for each model is reported in the legend.} 
    \label{fig:sed}
\end{figure}

Using our new ALMA observations in band 3 and band 6, we were able to detect an interacting companion and disentangle its emission from the one of the QSOs. Fig. \ref{fig:sed} shows the SED of the dust emission associated with the companion derived from the measurements in band 3 and band 6. We modelled it with a modified black body (MBB) function given as follows:
 \begin{equation}\label{eqsed}
     S_{\nu_{\rm obs}}^{\rm obs} = S_{\nu/(1+z)}^{\rm obs} = \dfrac{\Omega}{(1+z)^3}[B_{\nu}(T_{\rm dust}(z))-B_{\nu}(T_{\rm CMB}(z))](1-e^{-\tau_{\nu}}), 
 \end{equation}

\noindent where $\Omega = (1+z)^4A_{\rm gal}D_{\rm L}^{-2}$ is the solid angle as a function of the surface area of the galaxy, $A_{\rm gal}$, and the luminosity distance of the galaxy, $D_{\rm L}$. The dust optical depth is
\begin{equation}
    \tau_{\nu}=\dfrac{M_{\rm dust}}{A_{\rm gal}}k_0\biggl(\dfrac{\nu}{250\ \rm GHz}\biggr)^{\beta},
\end{equation}
\noindent with $\beta$ the emissivity index and $k_0 = 0.45\  \rm cm^{2}\ g^{-1}$ the mass absorption coefficient \citep{beelen+2006}. The solid angle was estimated using the size of the region where we extracted the flux of the companion, that is $\sim 4.0$ arcsec$^2$. The effect of the CMB on the dust temperature is given by
\begin{equation}
    T_{\rm dust}(z)=((T_{\rm dust})^{4+\beta}+T_0^{4+\beta}[(1+z)^{4+\beta}-1])^{\frac{1}{4+\beta}},
\end{equation}
\noindent with $T_0 = 2.73$ K.
We also accounted for the contribution of the CMB emission given by $B_{\nu}(T_{\rm CMB}(z)=T_0(1+z))$ \citep{dacunha2013}.

We considered one free parameter, $M_{\rm dust}$, and we fixed $T_{\rm dust}$ to 50 K, since $T_{\rm dust}$ cannot be constrained due to the lack of higher-frequency data and so we assumed a temperature similar to that found for the QSO \citep[$T_{\rm dust, QSO}=48\pm 2$ K, see][]{tripodi2023b}. We explored the one-dimensional parameter space using a Markov Chain Monte Carlo (MCMC) algorithm implemented in the \texttt{EMCEE} package \citep{foreman2013}, assuming a uniform prior for the fitting parameter: $10^{4} \ {\rm M_{\odot}}<M_{\rm dust}<10^{12}\ {\rm M_{\odot}}$. Considering $\beta$ to be in the range [1.0,3.0], we ran a MCMC with 20 chains and 1000 trials for each value of $\beta$ in that range with a step of 0.1. We find that only models with $2.0\leq \beta\leq 2.7$ are able to fit simultaneously the two points in band 6 and band 3, yielding $0.34 \times 10^7\ {\rm M_\odot} <M_{\rm dust}<1.3\times 10^7$ M$_\odot$, with burn-in phases of $\sim 20$ for each model. The best-fitting model with $T_{\rm dust}=50$ K, $\beta=2.0$, and $M_{\rm dust}=1.3\times 10^7$ M$_\odot$ is shown as a dashed cyan line in Fig. \ref{fig:sed}, and the one with $T_{\rm dust}=50$ K, $\beta=2.7$, and $M_{\rm dust}=3.4\times 10^6$ M$_\odot$ as a dotted cyan line. The recent ALMA 7m observation in band 9 did not detected any companion emission, probably due to the low resolution and sensitivity of the observation \citep{tripodi2023b}. However, we were able to derive a $3\sigma$ upper limit from the continuum map presented in \citet{tripodi2023b}, that is $2.6$ mJy. This upper limit is shown as a hollow green square and favours temperatures $\leq 50$ K (and $\beta>2.0$ for $T_{\rm dust}=50$ K). We considered a lower $T_{\rm dust}$ of 30 K, and we performed a similar MCMC fitting, yielding $2.4\leq \beta\leq 3.1$ and $0.7 \times 10^7\ {\rm M_\odot} <M_{\rm dust}<2.6\times 10^7$ M$_\odot$, with burn-in phases of $\sim 20$ for each model. The dashed and dotted blue lines are the best-fitting curves at $T_{\rm dust}=30$ K, $\beta=2.4$, $M_{\rm dust}=2.6\times 10^7$ M$_\odot$, and $T_{\rm dust}=30$ K, $\beta=3.1$, $M_{\rm dust}=6.7\times 10^6$ M$_\odot$, respectively. 

We also estimated the total infrared (TIR) luminosity of the four best-fitting models by integrating from $8$ to $1000\ \mu$m rest-frame, obtaining $L_{\rm TIR}$ in the range $[0.3-3.4]\times 10^{12}\ \rm L_{\odot}$. This implies an SFR in the range $[35-344]\ \rm M_{\odot}\ yr^{-1}$ \citep{kennicutt1998}, adopting a Chabrier initial mass function \citep{chabrier2003}. Assuming a Salpeter IMF would imply an SFR higher than a factor of 1.7. 

Table \ref{tab:qso-comp} lists the gas and dust properties of the QSO host and the companion.

\subsection{Outflow properties and energetics}
\label{sec:outflow}
\begin{figure*}
    \centering
    \includegraphics[width=0.8\linewidth]{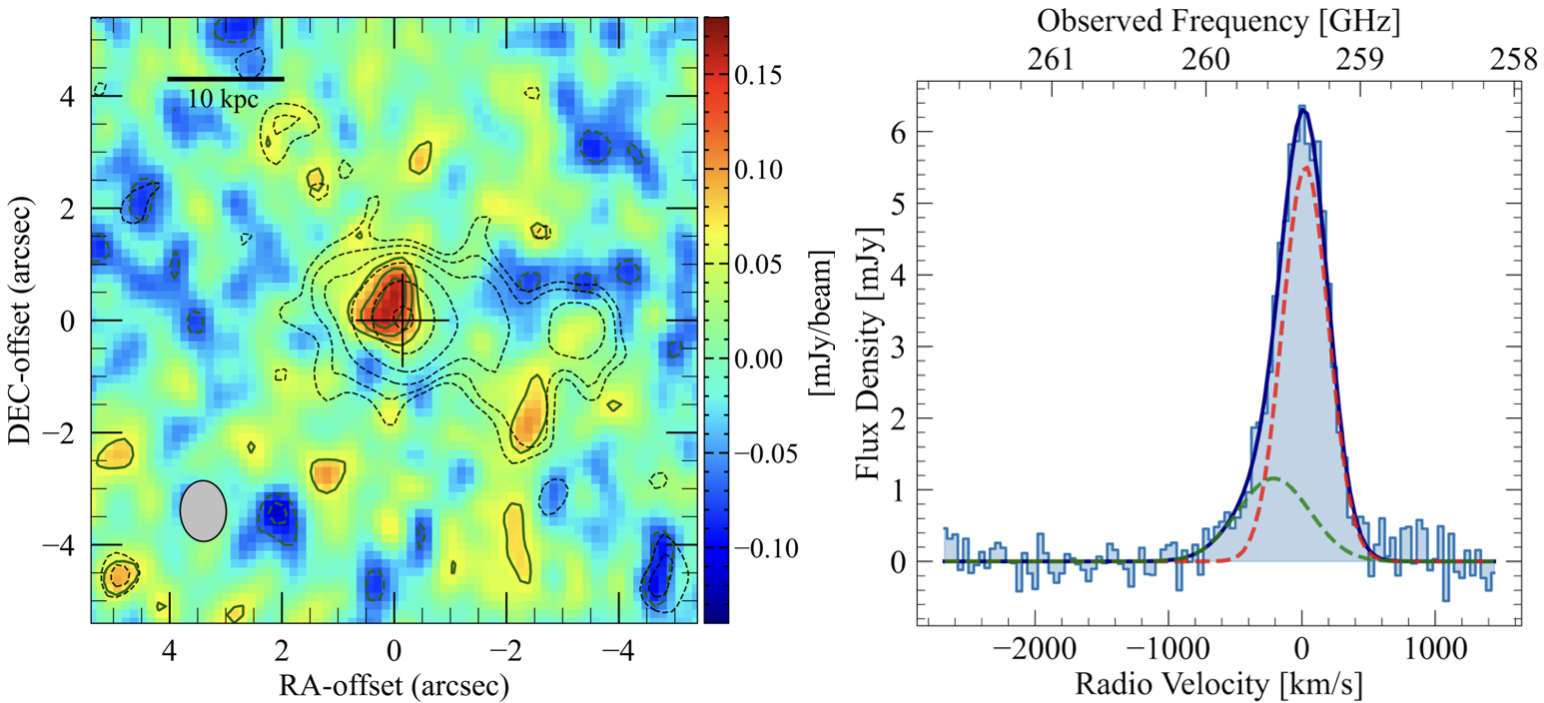}
    \caption{Outflow map and optimized [CII] spectrum. Left panel: [CII] outflow map made with v=[-500,-1000] km s$^{-1}$ with green contours at $-3,-2,2,3,\ {\rm and}\ 4\sigma$, with $\sigma=0.034$ mJy/beam, and dashed black contours from the full [CII] map, as in the top left panel of Fig. \ref{fig:1}. Right panel: the [CII] spectrum was extracted from a central region of 1.2 arcsec radius and binned at 40 km s$^{-1}$. The solid blue line is the total fit, composed of two Gaussian components (narrow, as a dashed red line; broad, as a dashed green line). The broad component is associated with an outflowing emission.}
    \label{fig:4}
\end{figure*}

The [CII] spectrum extracted from a circular region of $\sim 2$ arcsec radius shows emission towards high velocities (up to $\sim 1000$ km s$^{-1}$) on the blue side of the [CII] line. To ensure the highest S/N for this high-velocity emission, we optimized the aperture considering a central region of $\sim 1.2$ arcsec radius, from which we extracted the [CII] spectrum shown in the right panel of Fig. \ref{fig:4}. Firstly, we modelled the line with one single Gaussian, obtaining FWHM$= 467\pm 10$ km s$^{-1}$, and a peak frequency of 259.378 GHz, corresponding to $z=6.327$, which is in agreement with that found by \citet{wang2019}. We set the 0-velocity reference of the spectrum based on our peak frequency. As a first-order approach to estimate the flux and velocity of the high-velocity emission, we subtracted the Gaussian fit to the spectrum and found residual emission in [-800,-300] km s$^{-1}$. We integrated the spectrum in that range, finding an integrated flux of 0.155 Jy km s$^{-1}$. This corresponds to a luminosity of $1.64\times 10^8\ \rm L_{\odot}$. We computed the projected velocity, $v_{\rm 98}$, defined as the velocity at which the integrated flux of the high-velocity emission is 98\% of the total integrated flux with respect to the systemic velocity, finding $v_{\rm 98}$=720 km s$^{-1}$.

Secondly, we fitted the [CII] line with two Gaussian components in order to model the narrow and broader emission simultaneously, running the MCMC algorithm implemented in the \texttt{EMCEE} package \citep{foreman2013} with six free parameters: mean peak frequency, peak value, and FWHM for both the narrow and broad components. We adopted uniform priors for all the parameters, ensuring that the mean peak frequencies of the narrow and the broad components do not overlap. We were able to disentangle the two components, finding a FWHM$_{\rm N}= 407\pm 10$ km s$^{-1}$, a peak$_{\rm N} = 27\pm 5$ km s$^{-1}$, and an integrated flux, $(S{\rm d}v)_{\rm N}= 2.39\pm 0.11$ Jy km s$^{-1}$ for the narrow component (shown as a dashed red line), and a FWHM$_{\rm B}=635\pm 20$ km s$^{-1}$, a peak$_{\rm B}= -211\pm 57 $ km s$^{-1}$, and an integrated flux, $(S{\rm d}v)_{\rm B}= 0.78\pm 0.21$ Jy km s$^{-1}$ for the broad one (shown as a dashed green line). The total integrated flux of the line is indeed $(S{\rm d}v)_{\rm tot} = 3.17\pm 0.16$ Jy km s$^{-1}$, corresponding to a luminosity of $L_{\rm [CII]} = 3.36\times 10^9\ \rm L_{\odot}$. The left panel of Fig. \ref{fig:4} shows the  map of the high-velocity component of [CII], obtained by collapsing the velocity channels $[-500,-1000]$ km s$^{-1}$ from the continuum-subtracted cube\footnote{We selected the velocity range $[-500,-1000]$ km s$^{-1}$ in order not to be contaminated by the narrow component (that tends to zero at $\sim -500$ km s$^{-1}$), and since the broad component goes to zero at $\sim -1000$ km s$^{-1}$.} and overplotting the total [CII] emission map (dashed black line). Performing a 2D Gaussian fit we obtained a peak flux of $0.16\pm 0.02$ mJy/beam, a flux density of $0.30\pm 0.07$ mJy, and a FWHM size of $(0.95\pm 0.37)\times (0.71\pm 0.46)$ arcsec$^2$ (that corresponds to $\sim 5.4\times 4.0$ kpc$^2$), offset by 0.3 arcsec ($\sim 1.7$ kpc) towards the northeast, in other words in a direction almost perpendicular to the plane of the interacting system.

We interpret such a broad [CII] wing as an indication of a cold outflow, since velocities as high as 1000 km/s are unlikely to be associated with tidally stripped gas, especially given that, even more interestingly, it is elongated in the northeast direction similarly to the radio jet found by \citet{sbarrato2021}. This interpretation is better discussed in Sect. \ref{sec:disc}.

We computed the [CII] luminosity, $L_{\rm out}=8.27\times 10^8\ \rm L_{\odot}$ of the outflowing gas, and the upper and lower limits of the outflow radius, $R_{\rm min}=2.4$ kpc and $R_{\rm max}=5.5$ kpc, respectively.\footnote{The upper and lower outflow radii were derived by considering the uncertainties on the estimates of the FWHM size and the displacement between the peak of the QSO emission and of the outflow emission, that is $0.3$ arcsec$\sim 1.7$ kpc.}  Assuming a $L_{\rm [CII]}$ to gas mass conversion as in PDR regions, the corresponding neutral gas mass in PDR can be derived with the relation from \citet{Hailey2010} (see also \citealt{bischetti2019b}),

\begin{equation}
    \label{eq:cii_mass}
    \frac{M_{\rm HI}}{M_{\odot}} = 0.77\biggl(\frac{0.7L_{\rm [CII]}}{L_{\odot}}\biggr)\biggl(\frac{1.4\times 10^{-4}}{X_{C^+}}\biggr)\times\frac{1+2e^{-91{\rm K}/T}+n_{\rm crit}/n}{2e^{-91{\rm K}/T}}
\end{equation}

\noindent where $X_{C^+}$ is the [CII] fraction per hydrogen atom, T is the gas temperature, $n$ is the gas density, and $n_{\rm crit}\sim 3 \times 10^3$ cm$^{-3}$  is the [CII]$\lambda$158$\mu$m critical density. This yields $M_{\rm out}=1.1\times 10^9\ \rm M_{\odot}$. Assuming the scenario of time-averaged expelled shells or clumps \citep{rupke2005}, we computed the mass outflow rate,
\begin{equation}
    \dot{M}_{\rm out} = \frac{v_{\rm out}\times M_{\rm out}}{R_{\rm out}}
,\end{equation}

\noindent where $v_{\rm out}=|\Delta v_{\rm broad}|+{\rm FWHM}_{\rm B}/2=560$ km s$^{-1}$ and is $\Delta v_{\rm broad}$ is the velocity shift between the centroids of the narrow and broad components. Considering the upper and lower limits of the outflow radius, we have $118<\dot{M}_{\rm out}<269\ \rm M_\odot$ yr$^{-1}$. We also derived the kinetic power associated with the outflow as $\dot E_{\rm out} = \frac{1}{2} \dot M_{\rm out}\times v^2_{\rm out} =  (1.2-2.7)\times 10^{43} \rm erg ~s^{-1}$, and the wind momentum load as

\begin{equation}
    \frac{\dot P_{\rm out}}{\dot P_{\rm AGN}} = \frac{\dot M_{\rm out}\times v_{\rm out}}{L_{\rm bol}/c}=(0.01-0.02)
,\end{equation}
\noindent where $\dot P_{\rm AGN}$ is the AGN radiation momentum rate, and where we adopted the bolometric luminosity of $L_{\rm bol}= 4.29\times10^{14}\rm L_\odot$ estimated by \citet{wu2015} from the luminosity at $3000$ \AA.

The results for the spectrum fitting and the outflow energetics are reported in Table \ref{tab:lines}.

 We also estimated these quantities using the luminosity and $v_{\rm 98}$ from the first-order approach and the upper and lower limits of the radius, $R_{\rm min}=2.4$ kpc and $R_{\rm max}=5.5$ kpc, obtaining $30<\dot{M}_{\rm out}<68\ \rm M_\odot$ yr$^{-1}$,  $4.9<\dot E_{\rm out} < 11.3\times 10^{42} \rm erg ~s^{-1}$, and $0.002<\dot P_{\rm AGN}<0.006$. We consider these to be lower limits for the outflow energetics. Since the double Gaussian analysis is the approach commonly used in the literature to determine the outflow energetics at high-z \citep{maiolino2012, stanley2019, bischetti2019}, in the next section we compare the results derived by that approach (i.e., double Gaussian fit, see Tab. \ref{tab:lines}) with the results found in the literature for [CII] outflows in high-z QSOs and molecular outflows in low-z AGNs.

\begin{table}
\vspace{0.2cm}
		\caption{Properties of [CII] line and outflow energetics}
		\centering
		\begin{tabular}{lcc}
			\hline
            Quantity & Units & Value \\
			\hline
			$(S{\rm d}v)_{\rm N}$ & [Jy km s$^{-1}$] & $2.39 \pm 0.11$\\[0.05cm]
			FWHM$_{\rm N}$ & [km s$^{-1}$] &  $407 \pm 10$\\[0.05cm]
            peak$_{\rm N}$ & [km s$^{-1}$] & $27\pm 5$\\[0.05cm]
            $(S{\rm d}v)_{\rm B}$ & [Jy km s$^{-1}$] & $0.78 \pm 0.21$\\[0.05cm]
			FWHM$_{\rm B}$ & [km s$^{-1}$] &  $635 \pm 20$\\[0.05cm]
			peak$_{\rm B}$ & [km s$^{-1}$] & $-211\pm 57$\\ [0.05cm]
            $(S{\rm d}v)_{\rm tot}$ & [Jy km s$^{-1}$] & $3.17 \pm 0.16$\\[0.05cm]
            $L_{\rm [CII]}$ & [$\rm L_\odot$] & $3.36\times 10^9$\\[0.05cm]
			\grayline 
            $v_{\rm out}$ & [km s$^{-1}$] & 560\\[0.05cm]
            $\dot{M}_{\rm out}$ & [$\rm M_\odot\ yr^{-1}$] & 118-269\\[0.05cm]
            $\dot{E}_{\rm out}$ &[$10^{43}\ \rm erg\ s^{-1}$] & $(1.2-2.7)$\\[0.05cm]
            $\dot P_{\rm out}/\dot P_{\rm AGN}$ & & 0.01-0.02\\[0.05cm]
			
			\hline
		\end{tabular}
		\label{tab:lines}
		\flushleft 
		\footnotesize {{\bf Notes.} The subscript `N' refers to the narrow component, `B' to the broad component, and `tot' to the total [CII] emission line shown in Fig. \ref{fig:4}. $L_{\rm [CII]}$ was computed from the total integrated flux. The results for the outflow energetics are reported under the purple line.}
	\end{table}

\section{Discussion}
\label{sec:disc}

We presented results from new ALMA observations  of the [CII] emission line and of the continuum emission at 99.5 GHz and at 252.8 GHz of the QSO J0100+2802, which is the most luminous QSO at $z>6$. The high sensitivity of these observations enabled us to reveal the presence of an interacting companion and of a [CII] outflow. These features were undetected with higher-resolution observations \citep{neeleman2021}, reinforcing the idea that $\sim 1$ arcsec-resolution observations with a high sensitivity are ideal to probe the more extended and diffuse emission and to detect outflow signatures and tidally disrupted companions.

The [CII] emission, along with the continuum emissions in band 3 and band 6, show an elongated morphology that indicates the existence of an interactive companion. The [CII] and the band 6 continuum emissions are co-spatial and extend up to 30 kpc. We find a [CII] luminosity of $1.68\times 10^9\ \rm L_\odot$ for the companion, and of $4.70\times 10^9\ \rm L_\odot$ for the QSO host galaxy. These imply a neutral gas mass of $M_{\rm HI, comp}=2.30\times 10^9\ \rm M_\odot$ and $M_{\rm HI, host}=6.43\times 10^9\ \rm M_\odot$, respectively, by using Eq. \ref{eq:cii_mass}. We were also able to disentangle the QSO and the companion continuum emissions in band 3 and band 6, and from the fitting of the SED of the companion we obtained an SFR in the range $[35-344]\ \rm M_\odot\ yr^{-1}$, assuming $T_{\rm dust}=[30-50]$ K and $\beta=[2.0-3.1]$. The upper values for the companion SFR are indeed comparable or higher than the SFR found for the QSO host galaxy, which, along with the gas masses estimated above, suggests that both sources are gas-rich and that a major merger may be at the origin of the boosted star formation. 

The presence of a companion is consistent with the picture of QSOs living in overdense environments. In particular, a suite of simulations by \citet{dimascia2021} show that bright QSOs are part of complex, dust-rich merging systems, containing multiple sources (accreting black holes and/or star-forming galaxies). \citet{costa2019} also find a large number of satellite galaxies that will eventually fall on the central QSO host  at high-z in cosmological, radiation-hydrodynamic simulations. J0100+2802 is indeed found to have an exceptionally small proximity zone, given its extreme brightness, \citep{eilers2017, davies2020} and no detected Ly$\alpha$ nebula \citep{farina2019}. These are consistent with a scenario of a young black hole accretion episode and a recent merger, and also with the presence of a weak outflow, since the majority of the outflow emission would be intercepted by an over-abundance of neutral or dusty gas in the host galaxy \citep{costa2022}. Moreover, J0100+2802 is found to reside in a strong overdensity composed of 24 galaxies \citet{kashino2023}, detected in JWST-NIRCam slitless spectroscopy. However, thanks to the powerful capabilities of ALMA we were able to reveal this nearest interacting companion, which was undetected even by JWST-NIRCam imaging 
\citep{eilers2022}, probably because of both the overwhelming light from the QSO at optical and UV wavelengths and heavy dust obscuration, also supported by the substantial mass of dust detected at the position of the companion ($M_{\rm dust}\sim (0.3-2.6)\times 10^7\ \rm M_{\rm dust}$).

\begin{figure*}
    \centering
    \includegraphics[width=1\linewidth]{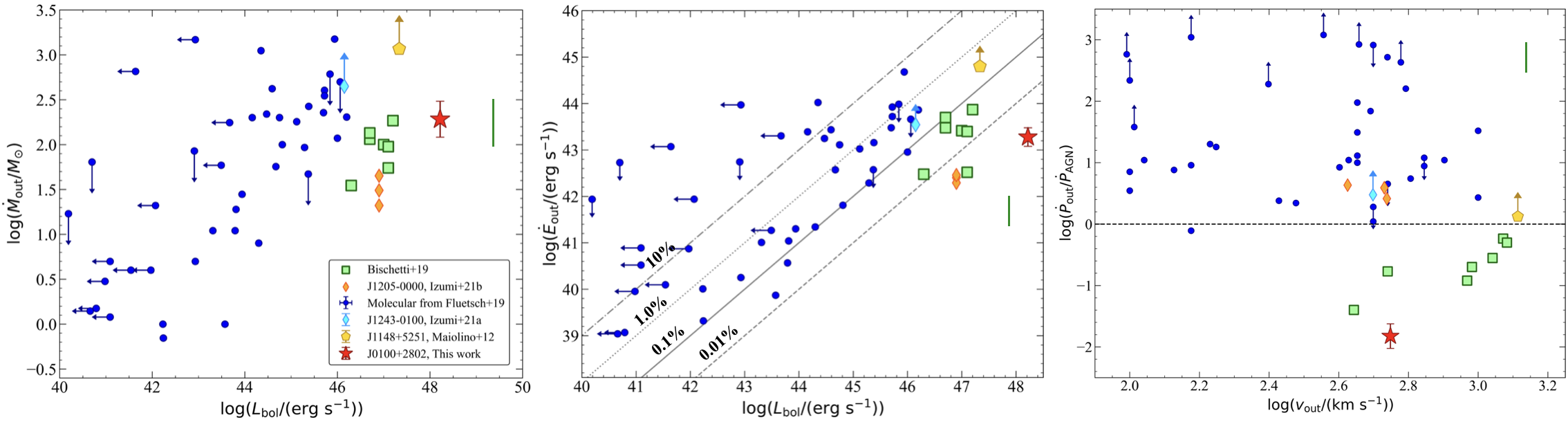}
    \caption{Energetics for the outflow in J0100+2802 (a red star) compared with different stacked spectra (hollow green squares, see legend; \citealt{bischetti2019}), [CII] outflows in QSO J1148+5251 at z=6.42 \citep[gold pentagon, ][]{maiolino2012} and in two QSOs at z=6.7 and z=7.07 belonging to the Subaru High-z Exploration of Low-luminosity Quasars (SHELLQs) sample \citep[diamonds,][]{izumi2021a,izumi2021b}, and molecular outflows in a sample of low-z AGNs \citep[blue dots,][]{fluetsch2019}.
    Left panel: mass outflow rate as a function of the bolometric luminosity. The typical $\sim 0.3$dex uncertainty on the outflow rate found in the sample of \citet{bischetti2019} is shown by the vertical green line.
    The error bars associated with the red stars mark the lower and upper limits found for J0100+2802, reported in Table \ref{tab:lines}. Central panel: Kinetic power as a function of the bolometric luminosity. The dot-dashed, dotted, solid, and dashed lines indicate kinetic powers that are 10\%, 1\%, 0.1\%, and 0.01\% of the bolometric luminosity. Right panel: momentum load factor as a function of the outflow velocity. The horizontal line corresponds to $\dot{P}_{\rm out}=\dot{P}_{\rm AGN}$.
    }
    \label{fig:outflow}
\end{figure*}

Studying the kinematics of the [CII] emission line, we found a velocity gradient oriented towards the companion position and a structure in the PV diagram that mimics a rotating disc. The high-velocity dispersion seen horizontally along the whole emission suggests the presence of a current or future merging process. The structure seen both in the PV diagram and in the momenta suggests two possible interpretations: (1) the arm at positive velocities in the PV may arise from the gas that is moving from the companion to the QSO host galaxy (called the “merging bridge”); (2) we might be witnessing the initial phases of the settling of a rotating gas structure during the merging process between the two sources. Indeed, similar kinematics has been found in a simulated merging system called Adenia in \citet{rizzo2022}. It is formed by two merging galaxies and presents a similar structure in the [CII] flux distribution, in the velocity map, and in the dispersion map (see Fig. 4 of \citet{rizzo2022}). However, the PV diagram of the Adenia galaxy is much more disturbed than the PV of J0100+2802. This may suggest the presence of some rotation entangled with the merging process, supporting a scenario in which we are witnessing the first phases of the settling of a rotating disc. However, the resolution of our observation did not allow us to distinguish between merging and rotation processes. \citet{neeleman2021} analysed a higher resolution ($\sim 0.2$ arcsec) and lower sensitivity observation of the [CII] of J0100+2802 and they did not find any sign of merger or any velocity gradient in the velocity map, which, on the contrary, seemed quite disturbed. This once again highlights the role of sensitivity in allowing a detailed analysis of the characteristics of the host galaxies.

We interpreted the broad component on the blue side of the [CII] line spectrum as an indication of an outflow. The outflow emission is resolved with a FWHM size of $\sim 5.4\times 4.0$ kpc$^2$, and it is located $\sim 0.3$ arcsec northeast of the QSO, in a similar direction to the radio jet found by \citet{sbarrato2021}, suggesting that the outflow is likely jet-driven. Both the outflow and the jet are aligned almost perpendicularly with the western extension (merger). 
Alternative interpretations of the broad wing are that it could be inflowing gas or a fainter companion. However, the alignment with the radio jet, together with the high dispersion found for the broad component (FWHM$_{\rm B}=635$ km s$^{-1}$), suggest that these interpretations are unlikely, and favour the outflow scenario. Indeed, the high dispersion would imply a very massive companion that however remains undetected in JWST observations \citep{kashino2023, eilers2022}. The presence of only a blue component of the wing may appear problematic in the context of simple bipolar outflow models. However, asymmetric outflows have already been observed \citep{fluetsch2019,bischetti2019}. Moreover, seeing the blueshifted [CII] component only can also be due to a high [CII] optical depth, which is often found in such gas-rich and compact galaxies \citep{papadopoulos2010, neri2014, gullberg2015}, and this would prevent us from seeing the redshifted component. We determined the mass outflow rate, $\dot{M}_{\rm out}= (118-269)\ \rm M_\odot\ yr^{-1}$, by considering the upper and lower limits of the outflow radius. The average mass outflow rate is comparable to the SFRs found for both the QSO host galaxy and the companion. This may support a scenario in which the outflow is mostly driven by SF. Moreover, in simulations, QSO companions that are directly impacted by the outflow are found to have their SFR increased by a factor of two to three, and tend to be more massive \citep{zana2022}. Considering the SFR of the QSO host, the mass outflow rate, and the molecular gas mass of $5.4\times 10^9\ \rm M_\odot$ found by \citet{wang2019}, we infer a depletion time of 10-13 Myr, which seems to imply a rapid quenching of the QSO host galaxy. However, given that the outflow velocity is $\sim$560 km/s, the evacuated gas may not escape the halo and may eventually fall back on the host galaxy, triggering another phase of galaxy growth. Indeed the host galaxy of J0100+2802 is expected to experience massive growth, since this QSO is already strongly off the local relation in the $M_{\rm BH}-M_{\rm dyn}$ plane (see Fig. 3 of \citealt{tripodi2023b}), given the high BH mass. 

We find the kinetic power to be $\dot{E}_{\rm out}= (1.2-2.7)\times 10^{43}\ \rm erg\ s^{-1}$ and the momentum load, $\dot P_{\rm out}/\dot P_{\rm AGN}= 0.01-0.02$, considering the upper and lower limits of the outflow radius. 
In Fig. \ref{fig:outflow}, we compare our results (red stars) with those of [CII] outflows in QSOs at $4.5<z<7.1$ from spectra stacking \citep{bischetti2019} of [CII] outflows in QSO SDSS J1148+5251 \citep{maiolino2012, cicone2015}, in QSO HSC J1205-0000\footnote{Note that \citet{izumi2021b} also propose a merger as a possible interpretation for the broad component in J1205-0000.}, in QSO HSC J1243+0100 at z=7.07 belonging to the Subaru High-z Exploration of Low-luminosity Quasars (SHELLQs) sample \citep{izumi2021a,izumi2021b}, and of molecular outflows in a sample of low-z AGNs \citep{fluetsch2019}.
We find that the properties of our [CII] outflow are at the extreme end of the population for the mass outflow rate and the kinetic power. This is expected, given the high luminosity of this QSO. Even though it is challenging to directly compare different phases of outflowing gas, we acknowledge that the energy of [CII] outflows detected in high-z QSOs is systematically lower than that of local molecular outflows, which show -- on average -- an increasing trend in terms of the mass outflow rate and kinetic power with bolometric luminosity and a high momentum load. In our case, the low momentum load factor suggests that the outflow is either energy-driven but with poor coupling with the host galaxy ISM, or is driven by direct radiation pressure onto the dusty clouds \citep[e.g.,][]{ishibashi2018, bischetti2019, bischetti2019b}. Either cases support the fact that the outflow cannot be very effective at removing gas from the entire galaxy \citep{gabor2014, costa2015, costa2018, bourne2015, roos2015, valentini2021}. 

\section{Summary}
\label{sec:summ}

In this work, new ALMA observations of the [CII] emission line and of the continuum emission in band 6 and band 3 of the HYPERION QSO J0100+2802 reveal an interesting new picture of the most luminous QSO at $z>6$. We find an interactive companion and a high-velocity cold outflow that were undetected by previous higher-resolution observations \citep{neeleman2021} and by JWST-NIRCam imaging \citep{eilers2022}. The [CII] emission and its underlying continuum are horizontally elongated with multiple peaks, and the [CII] channel maps show, even more clearly, a clumpy morphology and a tidal tail. These, together with the enhanced [CII] velocity dispersion along the direction of the elongation, support the scenario in which the companion is merging with the QSO-host. We derived a range for the SFR of the companion given also the upper limit in band 9, yielding an SFR of $\sim [35-344]\ \rm M_\odot yr^{-1}$. From the analysis of the outflow in the [CII] spectrum, we obtain a mass outflow rate of  $\dot{M}_{\rm out}= (118-269)\ \rm M_\odot\ yr^{-1}$, considering that the outflow emission is resolved with a size of $\sim 5.4 \times 4.0$ kpc$^{2}$. Computing the outflow energetics, we conclude that the outflow cannot be very effecting at removing the gas from the galaxy.

These results stress the importance of deep low-resolution ALMA observations for the study of QSOs at the EoR. It is now necessary to follow up on this interesting system with deep high-resolution observations so as to accurately determine the kinematics of both the QSO host and the companion.

\vspace{1cm}
\noindent \textit{Acknowledgments.} We thank F. Lelli and D. Kashino for the interesting and helpful discussions. We are grateful to C. Cicone for their comments during the preparation of the proposal. This paper makes use of the following ALMA data: ADS/JAO.ALMA\#2021.2.00151.S. ALMA is a partnership of ESO (representing its member states), NFS (USA) and NINS (Japan), together with NRC (Canada), MOST and ASIAA (Taiwan) and KASI (Republic of Korea), in cooperation with the Republic of Chile. The Joint ALMA Observatory is operated by ESO, AUI/NRAO and NAOJ. The project leading to this publication has received support from ORP, that is funded by the European Union’s Horizon 2020 research
and innovation programme under grant agreement No 101004719 [ORP]. RT acknowledges financial support from the University of Trieste. RT, CF, FF, MB, EP acknowledge support from PRIN MIUR project “Black Hole winds and the Baryon Life Cycle of Galaxies: the stone-guest at the galaxy evolution supper”, contract \#2017PH3WAT. EP, LZ and CF acknowledge financial support from the Bando Ricerca Fondamentale INAF 2022 Large Grant "Toward an holistic view of the Titans: multi-band observations of $z>6$ QSOs powered by greedy supermassive black-holes". RM and JS acknowledges ERC Advanced Grant 695671 QUENCH, and support from the UK Science and Technology Facilities Council (STFC). RM also acknowledges funding from a research professorship from the Royal Society. SC acknowledges support from the European Union (ERC, WINGS,101040227). 
    \textit{Facilities:} ALMA. \textit{Software:} astropy \citep{astropy2022}, Matplotlib \citep{matplotlib2007}, SciPy \citep{scipy2023}, CASA (v5.1.1-5, \citealt{casa2022}).

\vspace{5mm}
          
\bibliography{biblio}{}

\begin{thebibliography}{79}
\expandafter\ifx\csname natexlab\endcsname\relax\def\natexlab#1{#1}\fi

\bibitem[{{Astropy Collaboration} {et~al.}(2022){Astropy Collaboration},
  {Price-Whelan}, {Lim}, {Earl}, {Starkman}, {Bradley}, {Shupe}, {Patil},
  {Corrales}, {Brasseur}, {N{\"o}the}, {Donath}, {Tollerud}, {Morris},
  {Ginsburg}, {Vaher}, {Weaver}, {Tocknell}, {Jamieson}, {van Kerkwijk},
  {Robitaille}, {Merry}, {Bachetti}, {G{\"u}nther}, {Aldcroft},
  {Alvarado-Montes}, {Archibald}, {B{\'o}di}, {Bapat}, {Barentsen},
  {Baz{\'a}n}, {Biswas}, {Boquien}, {Burke}, {Cara}, {Cara}, {Conroy},
  {Conseil}, {Craig}, {Cross}, {Cruz}, {D'Eugenio}, {Dencheva}, {Devillepoix},
  {Dietrich}, {Eigenbrot}, {Erben}, {Ferreira}, {Foreman-Mackey}, {Fox},
  {Freij}, {Garg}, {Geda}, {Glattly}, {Gondhalekar}, {Gordon}, {Grant},
  {Greenfield}, {Groener}, {Guest}, {Gurovich}, {Handberg}, {Hart},
  {Hatfield-Dodds}, {Homeier}, {Hosseinzadeh}, {Jenness}, {Jones}, {Joseph},
  {Kalmbach}, {Karamehmetoglu}, {Ka{\l}uszy{\'n}ski}, {Kelley}, {Kern},
  {Kerzendorf}, {Koch}, {Kulumani}, {Lee}, {Ly}, {Ma}, {MacBride}, {Maljaars},
  {Muna}, {Murphy}, {Norman}, {O'Steen}, {Oman}, {Pacifici}, {Pascual},
  {Pascual-Granado}, {Patil}, {Perren}, {Pickering}, {Rastogi}, {Roulston},
  {Ryan}, {Rykoff}, {Sabater}, {Sakurikar}, {Salgado}, {Sanghi}, {Saunders},
  {Savchenko}, {Schwardt}, {Seifert-Eckert}, {Shih}, {Jain}, {Shukla}, {Sick},
  {Simpson}, {Singanamalla}, {Singer}, {Singhal}, {Sinha}, {Sip{\H{o}}cz},
  {Spitler}, {Stansby}, {Streicher}, {{\v{S}}umak}, {Swinbank}, {Taranu},
  {Tewary}, {Tremblay}, {de Val-Borro}, {Van Kooten}, {Vasovi{\'c}}, {Verma},
  {de Miranda Cardoso}, {Williams}, {Wilson}, {Winkel}, {Wood-Vasey}, {Xue},
  {Yoachim}, {Zhang}, {Zonca}, \& {Astropy Project Contributors}}]{astropy2022}
{Astropy Collaboration}, {Price-Whelan}, A.~M., {Lim}, P.~L., {et~al.} 2022,
  \apj, 935, 167

\bibitem[{{Beelen} {et~al.}(2006){Beelen}, {Cox}, {Benford}, {Dowell},
  {Kov{\'a}cs}, {Bertoldi}, {Omont}, \& {Carilli}}]{beelen+2006}
{Beelen}, A., {Cox}, P., {Benford}, D.~J., {et~al.} 2006, \apj, 642, 694

\bibitem[{{Bischetti} {et~al.}(2019{\natexlab{a}}){Bischetti}, {Maiolino},
  {Carniani}, {Fiore}, {Piconcelli}, \& {Fluetsch}}]{bischetti2019}
{Bischetti}, M., {Maiolino}, R., {Carniani}, S., {et~al.} 2019{\natexlab{a}},
  \aap, 630, A59

\bibitem[{{Bischetti} {et~al.}(2019{\natexlab{b}}){Bischetti}, {Piconcelli},
  {Feruglio}, {Fiore}, {Carniani}, {Brusa}, {Cicone}, {Vignali}, {Bongiorno},
  {Cresci}, {Mainieri}, {Maiolino}, {Marconi}, {Nardini}, \&
  {Zappacosta}}]{bischetti2019b}
{Bischetti}, M., {Piconcelli}, E., {Feruglio}, C., {et~al.} 2019{\natexlab{b}},
  \aap, 628, A118

\bibitem[{{Bourne} {et~al.}(2015){Bourne}, {Zubovas}, \&
  {Nayakshin}}]{bourne2015}
{Bourne}, M.~A., {Zubovas}, K., \& {Nayakshin}, S. 2015, \mnras, 453, 1829

\bibitem[{{Brusa} {et~al.}(2018){Brusa}, {Cresci}, {Daddi}, {Paladino},
  {Perna}, {Bongiorno}, {Lusso}, {Sargent}, {Casasola}, {Feruglio},
  {Fraternali}, {Georgiev}, {Mainieri}, {Carniani}, {Comastri}, {Duras},
  {Fiore}, {Mannucci}, {Marconi}, {Piconcelli}, {Zamorani}, {Gilli}, {La
  Franca}, {Lanzuisi}, {Lutz}, {Santini}, {Scoville}, {Vignali}, {Vito},
  {Rabien}, {Busoni}, \& {Bonaglia}}]{brusa2018}
{Brusa}, M., {Cresci}, G., {Daddi}, E., {et~al.} 2018, \aap, 612, A29

\bibitem[{{Carniani} {et~al.}(2017){Carniani}, {Maiolino}, {Pallottini},
  {Vallini}, {Pentericci}, {Ferrara}, {Castellano}, {Vanzella}, {Grazian},
  {Gallerani}, {Santini}, {Wagg}, \& {Fontana}}]{carniani2017}
{Carniani}, S., {Maiolino}, R., {Pallottini}, A., {et~al.} 2017, \aap, 605, A42

\bibitem[{{CASA Team} {et~al.}(2022){CASA Team}, {Bean}, {Bhatnagar}, {Castro},
  {Donovan Meyer}, {Emonts}, {Garcia}, {Garwood}, {Golap}, {Gonzalez Villalba},
  {Harris}, {Hayashi}, {Hoskins}, {Hsieh}, {Jagannathan}, {Kawasaki},
  {Keimpema}, {Kettenis}, {Lopez}, {Marvil}, {Masters}, {McNichols},
  {Mehringer}, {Miel}, {Moellenbrock}, {Montesino}, {Nakazato}, {Ott}, {Petry},
  {Pokorny}, {Raba}, {Rau}, {Schiebel}, {Schweighart}, {Sekhar}, {Shimada},
  {Small}, {Steeb}, {Sugimoto}, {Suoranta}, {Tsutsumi}, {van Bemmel},
  {Verkouter}, {Wells}, {Xiong}, {Szomoru}, {Griffith}, {Glendenning}, \&
  {Kern}}]{casa2022}
{CASA Team}, {Bean}, B., {Bhatnagar}, S., {et~al.} 2022, \pasp, 134, 114501

\bibitem[{{Chabrier}(2003)}]{chabrier2003}
{Chabrier}, G. 2003, \pasp, 115, 763

\bibitem[{{Cicone} {et~al.}(2015){Cicone}, {Maiolino}, {Gallerani}, {Neri},
  {Ferrara}, {Sturm}, {Fiore}, {Piconcelli}, \& {Feruglio}}]{cicone2015}
{Cicone}, C., {Maiolino}, R., {Gallerani}, S., {et~al.} 2015, \aap, 574, A14

\bibitem[{{Cormier} {et~al.}(2015){Cormier}, {Madden}, {Lebouteiller}, {Abel},
  {Hony}, {Galliano}, {R{\'e}my-Ruyer}, {Bigiel}, {Baes}, {Boselli},
  {Chevance}, {Cooray}, {De Looze}, {Doublier}, {Galametz}, {Hughes},
  {Karczewski}, {Lee}, {Lu}, \& {Spinoglio}}]{cormier2015}
{Cormier}, D., {Madden}, S.~C., {Lebouteiller}, V., {et~al.} 2015, \aap, 578,
  A53

\bibitem[{{Costa} {et~al.}(2022){Costa}, {Arrigoni Battaia}, {Farina},
  {Keating}, {Rosdahl}, \& {Kimm}}]{costa2022}
{Costa}, T., {Arrigoni Battaia}, F., {Farina}, E.~P., {et~al.} 2022, \mnras,
  517, 1767

\bibitem[{{Costa} {et~al.}(2019){Costa}, {Rosdahl}, \& {Kimm}}]{costa2019}
{Costa}, T., {Rosdahl}, J., \& {Kimm}, T. 2019, \mnras, 489, 5181

\bibitem[{{Costa} {et~al.}(2018){Costa}, {Rosdahl}, {Sijacki}, \&
  {Haehnelt}}]{costa2018}
{Costa}, T., {Rosdahl}, J., {Sijacki}, D., \& {Haehnelt}, M.~G. 2018, \mnras,
  479, 2079

\bibitem[{{Costa} {et~al.}(2015){Costa}, {Sijacki}, \& {Haehnelt}}]{costa2015}
{Costa}, T., {Sijacki}, D., \& {Haehnelt}, M.~G. 2015, \mnras, 448, L30

\bibitem[{{da Cunha} {et~al.}(2013){da Cunha}, {Groves}, {Walter}, {Decarli},
  {Weiss}, {Bertoldi}, {Carilli}, {Daddi}, {Elbaz}, {Ivison}, {Maiolino},
  {Riechers}, {Rix}, {Sargent}, \& {Smail}}]{dacunha2013}
{da Cunha}, E., {Groves}, B., {Walter}, F., {et~al.} 2013, \apj, 766, 13

\bibitem[{{Davies} {et~al.}(2020){Davies}, {Wang}, {Eilers}, \&
  {Hennawi}}]{davies2020}
{Davies}, F.~B., {Wang}, F., {Eilers}, A.-C., \& {Hennawi}, J.~F. 2020, \apjl,
  904, L32

\bibitem[{{Davies} {et~al.}(2011){Davies}, {F{\"o}rster Schreiber}, {Cresci},
  {Genzel}, {Bouch{\'e}}, {Burkert}, {Buschkamp}, {Genel}, {Hicks}, {Kurk},
  {Lutz}, {Newman}, {Shapiro}, {Sternberg}, {Tacconi}, \& {Wuyts}}]{davies2011}
{Davies}, R., {F{\"o}rster Schreiber}, N.~M., {Cresci}, G., {et~al.} 2011,
  \apj, 741, 69

\bibitem[{{Di Mascia} {et~al.}(2021){Di Mascia}, {Gallerani}, {Behrens},
  {Pallottini}, {Carniani}, {Ferrara}, {Barai}, {Vito}, \&
  {Zana}}]{dimascia2021}
{Di Mascia}, F., {Gallerani}, S., {Behrens}, C., {et~al.} 2021, \mnras, 503,
  2349

\bibitem[{{Di Matteo} {et~al.}(2005){Di Matteo}, {Springel}, \&
  {Hernquist}}]{dimatteo2005}
{Di Matteo}, T., {Springel}, V., \& {Hernquist}, L. 2005, \nat, 433, 604

\bibitem[{{Eilers} {et~al.}(2017){Eilers}, {Davies}, {Hennawi}, {Prochaska},
  {Luki{\'c}}, \& {Mazzucchelli}}]{eilers2017}
{Eilers}, A.-C., {Davies}, F.~B., {Hennawi}, J.~F., {et~al.} 2017, \apj, 840,
  24

\bibitem[{{Eilers} {et~al.}(2023){Eilers}, {Simcoe}, {Yue}, {Mackenzie},
  {Matthee}, {{\v{D}}urov{\v{c}}{\'\i}kov{\'a}}, {Kashino}, {Bordoloi}, \&
  {Lilly}}]{eilers2022}
{Eilers}, A.-C., {Simcoe}, R.~A., {Yue}, M., {et~al.} 2023, \apj, 950, 68

\bibitem[{{Fan} {et~al.}(2023){Fan}, {Ba{\~n}ados}, \& {Simcoe}}]{fan2022}
{Fan}, X., {Ba{\~n}ados}, E., \& {Simcoe}, R.~A. 2023, \araa, 61, 373

\bibitem[{{Farina} {et~al.}(2019){Farina}, {Arrigoni-Battaia}, {Costa},
  {Walter}, {Hennawi}, {Drake}, {Decarli}, {Gutcke}, {Mazzucchelli},
  {Neeleman}, {Georgiev}, {Eilers}, {Davies}, {Ba{\~n}ados}, {Fan}, {Onoue},
  {Schindler}, {Venemans}, {Wang}, {Yang}, {Rabien}, \& {Busoni}}]{farina2019}
{Farina}, E.~P., {Arrigoni-Battaia}, F., {Costa}, T., {et~al.} 2019, \apj, 887,
  196

\bibitem[{{Feruglio} {et~al.}(2017){Feruglio}, {Ferrara}, {Bischetti},
  {Downes}, {Neri}, {Ceccarelli}, {Cicone}, {Fiore}, {Gallerani}, {Maiolino},
  {Menci}, {Piconcelli}, {Vietri}, {Vignali}, \& {Zappacosta}}]{feruglio2017}
{Feruglio}, C., {Ferrara}, A., {Bischetti}, M., {et~al.} 2017, \aap, 608, A30

\bibitem[{{Fiore} {et~al.}(2017){Fiore}, {Feruglio}, {Shankar}, {Bischetti},
  {Bongiorno}, {Brusa}, {Carniani}, {Cicone}, {Duras}, {Lamastra}, {Mainieri},
  {Marconi}, {Menci}, {Maiolino}, {Piconcelli}, {Vietri}, \&
  {Zappacosta}}]{fiore2017}
{Fiore}, F., {Feruglio}, C., {Shankar}, F., {et~al.} 2017, \aap, 601, A143

\bibitem[{{Fluetsch} {et~al.}(2019){Fluetsch}, {Maiolino}, {Carniani},
  {Marconi}, {Cicone}, {Bourne}, {Costa}, {Fabian}, {Ishibashi}, \&
  {Venturi}}]{fluetsch2019}
{Fluetsch}, A., {Maiolino}, R., {Carniani}, S., {et~al.} 2019, \mnras, 483,
  4586

\bibitem[{{Foreman-Mackey} {et~al.}(2013){Foreman-Mackey}, {Hogg}, {Lang}, \&
  {Goodman}}]{foreman2013}
{Foreman-Mackey}, D., {Hogg}, D.~W., {Lang}, D., \& {Goodman}, J. 2013, \pasp,
  125, 306

\bibitem[{{Fujimoto} {et~al.}(2020){Fujimoto}, {Oguri}, {Nagao}, {Izumi}, \&
  {Ouchi}}]{fujimoto2020}
{Fujimoto}, S., {Oguri}, M., {Nagao}, T., {Izumi}, T., \& {Ouchi}, M. 2020,
  \apj, 891, 64

\bibitem[{{Gabor} \& {Bournaud}(2014)}]{gabor2014}
{Gabor}, J.~M. \& {Bournaud}, F. 2014, \mnras, 441, 1615

\bibitem[{{Gommers} {et~al.}(2023){Gommers}, {Virtanen}, {Burovski},
  {Haberland}, {Weckesser}, {Oliphant}, {Reddy}, {Cournapeau}, {Alexbrc},
  {Nelson}, {Peterson}, {Wilson}, {Roy}, {Endolith}, {Polat}, {Mayorov}, {Van
  Der Walt}, {Brett}, {Laxalde}, {Larson}, {Millman}, {Sakai}, {Lars},
  {Peterbell10}, {Van Mulbregt}, {Carey}, {Eric-Jones}, {McKibben}, {Kai}, \&
  {Kern}}]{scipy2023}
{Gommers}, R., {Virtanen}, P., {Burovski}, E., {et~al.} 2023, {scipy/scipy:
  SciPy 1.10.1}, Zenodo

\bibitem[{{Greene} {et~al.}(2020){Greene}, {Strader}, \& {Ho}}]{greene2020}
{Greene}, J.~E., {Strader}, J., \& {Ho}, L.~C. 2020, \araa, 58, 257

\bibitem[{{Gullberg} {et~al.}(2015){Gullberg}, {De Breuck}, {Vieira},
  {Wei{\ss}}, {Aguirre}, {Aravena}, {B{\'e}thermin}, {Bradford}, {Bothwell},
  {Carlstrom}, {Chapman}, {Fassnacht}, {Gonzalez}, {Greve}, {Hezaveh},
  {Holzapfel}, {Husband}, {Ma}, {Malkan}, {Marrone}, {Menten}, {Murphy},
  {Reichardt}, {Spilker}, {Stark}, {Strandet}, \& {Welikala}}]{gullberg2015}
{Gullberg}, B., {De Breuck}, C., {Vieira}, J.~D., {et~al.} 2015, \mnras, 449,
  2883

\bibitem[{{Hailey-Dunsheath} {et~al.}(2010){Hailey-Dunsheath}, {Nikola},
  {Stacey}, {Oberst}, {Parshley}, {Benford}, {Staguhn}, \&
  {Tucker}}]{Hailey2010}
{Hailey-Dunsheath}, S., {Nikola}, T., {Stacey}, G.~J., {et~al.} 2010, \apjl,
  714, L162

\bibitem[{{Harrison} {et~al.}(2018){Harrison}, {Costa}, {Tadhunter},
  {Fl{\"u}tsch}, {Kakkad}, {Perna}, \& {Vietri}}]{harrison2018}
{Harrison}, C.~M., {Costa}, T., {Tadhunter}, C.~N., {et~al.} 2018, Nature
  Astronomy, 2, 198

\bibitem[{{Hollenbach} \& {Tielens}(1999)}]{hollenbach1999}
{Hollenbach}, D.~J. \& {Tielens}, A.~G.~G.~M. 1999, Reviews of Modern Physics,
  71, 173

\bibitem[{{Hopkins} {et~al.}(2008){Hopkins}, {Hernquist}, {Cox}, \&
  {Kere{\v{s}}}}]{hopkins2008}
{Hopkins}, P.~F., {Hernquist}, L., {Cox}, T.~J., \& {Kere{\v{s}}}, D. 2008,
  \apjs, 175, 356

\bibitem[{{Hunter}(2007)}]{matplotlib2007}
{Hunter}, J.~D. 2007, Computing in Science and Engineering, 9, 90

\bibitem[{{Inayoshi} {et~al.}(2020){Inayoshi}, {Visbal}, \&
  {Haiman}}]{inayoshi2020}
{Inayoshi}, K., {Visbal}, E., \& {Haiman}, Z. 2020, \araa, 58, 27

\bibitem[{{Ishibashi} {et~al.}(2018){Ishibashi}, {Fabian}, \&
  {Maiolino}}]{ishibashi2018}
{Ishibashi}, W., {Fabian}, A.~C., \& {Maiolino}, R. 2018, \mnras, 476, 512

\bibitem[{{Izumi} {et~al.}(2021{\natexlab{a}}){Izumi}, {Matsuoka}, {Fujimoto},
  {Onoue}, {Strauss}, {Umehata}, {Imanishi}, {Kohno}, {Kawaguchi}, {Kawamuro},
  {Baba}, {Nagao}, {Toba}, {Inayoshi}, {Silverman}, {Inoue}, {Ikarashi},
  {Iwasawa}, {Kashikawa}, {Hashimoto}, {Nakanishi}, {Ueda}, {Schramm}, {Lee},
  \& {Suh}}]{izumi2021a}
{Izumi}, T., {Matsuoka}, Y., {Fujimoto}, S., {et~al.} 2021{\natexlab{a}}, \apj,
  914, 36

\bibitem[{{Izumi} {et~al.}(2021{\natexlab{b}}){Izumi}, {Onoue}, {Matsuoka},
  {Strauss}, {Fujimoto}, {Umehata}, {Imanishi}, {Kawamuro}, {Nagao}, {Toba},
  {Kohno}, {Kashikawa}, {Inayoshi}, {Kawaguchi}, {Iwasawa}, {Inoue}, {Goto},
  {Baba}, {Schramm}, {Suh}, {Harikane}, {Ueda}, {Silverman}, {Hashimoto},
  {Hashimoto}, {Ikarashi}, {Iono}, {Lee}, {Lee}, {Minezaki}, {Nakanishi},
  {Nakano}, {Tamura}, \& {Tang}}]{izumi2021b}
{Izumi}, T., {Onoue}, M., {Matsuoka}, Y., {et~al.} 2021{\natexlab{b}}, \apj,
  908, 235

\bibitem[{{Johnson} \& {Haardt}(2016)}]{johnson2016}
{Johnson}, J.~L. \& {Haardt}, F. 2016, \pasa, 33, e007

\bibitem[{{Kashino} {et~al.}(2023){Kashino}, {Lilly}, {Matthee}, {Eilers},
  {Mackenzie}, {Bordoloi}, \& {Simcoe}}]{kashino2023}
{Kashino}, D., {Lilly}, S.~J., {Matthee}, J., {et~al.} 2023, \apj, 950, 66

\bibitem[{{Kennicutt}(1998)}]{kennicutt1998}
{Kennicutt}, Robert~C., J. 1998, \araa, 36, 189

\bibitem[{Kormendy \& Ho(2013)}]{kormendy2013}
Kormendy, J. \& Ho, L.~C. 2013, Annual Review of Astronomy and Astrophysics,
  51, 511

\bibitem[{{Kroupa} {et~al.}(2020){Kroupa}, {Subr}, {Jerabkova}, \&
  {Wang}}]{kroupa2020}
{Kroupa}, P., {Subr}, L., {Jerabkova}, T., \& {Wang}, L. 2020, \mnras, 498,
  5652

\bibitem[{{Lelli} {et~al.}(2021){Lelli}, {Di Teodoro}, {Fraternali}, {Man},
  {Zhang}, {De Breuck}, {Davis}, \& {Maiolino}}]{lelli2021}
{Lelli}, F., {Di Teodoro}, E.~M., {Fraternali}, F., {et~al.} 2021, Science,
  371, 713

\bibitem[{{Lupi} {et~al.}(2021){Lupi}, {Haiman}, \& {Volonteri}}]{lupi2021}
{Lupi}, A., {Haiman}, Z., \& {Volonteri}, M. 2021, \mnras, 503, 5046

\bibitem[{{Maiolino} {et~al.}(2012){Maiolino}, {Gallerani}, {Neri}, {Cicone},
  {Ferrara}, {Genzel}, {Lutz}, {Sturm}, {Tacconi}, {Walter}, {Feruglio},
  {Fiore}, \& {Piconcelli}}]{maiolino2012}
{Maiolino}, R., {Gallerani}, S., {Neri}, R., {et~al.} 2012, \mnras, 425, L66

\bibitem[{{Meyer} {et~al.}(2022){Meyer}, {Walter}, {Cicone}, {Cox}, {Decarli},
  {Neri}, {Novak}, {Pensabene}, {Riechers}, \& {Weiss}}]{meyer2022}
{Meyer}, R.~A., {Walter}, F., {Cicone}, C., {et~al.} 2022, \apj, 927, 152

\bibitem[{{Neeleman} {et~al.}(2021){Neeleman}, {Novak}, {Venemans}, {Walter},
  {Decarli}, {Kaasinen}, {Schindler}, {Ba{\~n}ados}, {Carilli}, {Drake}, {Fan},
  \& {Rix}}]{neeleman2021}
{Neeleman}, M., {Novak}, M., {Venemans}, B.~P., {et~al.} 2021, \apj, 911, 141

\bibitem[{{Neri} {et~al.}(2014){Neri}, {Downes}, {Cox}, \& {Walter}}]{neri2014}
{Neri}, R., {Downes}, D., {Cox}, P., \& {Walter}, F. 2014, \aap, 562, A35

\bibitem[{{Novak} {et~al.}(2020){Novak}, {Venemans}, {Walter}, {Neeleman},
  {Kaasinen}, {Liang}, {Feldmann}, {Ba{\~n}ados}, {Carilli}, {Decarli},
  {Drake}, {Fan}, {Farina}, {Mazzucchelli}, {Rix}, \& {Wang}}]{novak2020}
{Novak}, M., {Venemans}, B.~P., {Walter}, F., {et~al.} 2020, \apj, 904, 131

\bibitem[{{Olsen} {et~al.}(2018){Olsen}, {Pallottini}, {Wofford}, {Chatzikos},
  {Revalski}, {Guzm{\'a}n}, {Popping}, {V{\'a}zquez-Semadeni}, {Magdis},
  {Richardson}, {Hirschmann}, \& {Gray}}]{olsen2018}
{Olsen}, K., {Pallottini}, A., {Wofford}, A., {et~al.} 2018, Galaxies, 6, 100

\bibitem[{{Papadopoulos} {et~al.}(2010){Papadopoulos}, {van der Werf}, {Isaak},
  \& {Xilouris}}]{papadopoulos2010}
{Papadopoulos}, P.~P., {van der Werf}, P., {Isaak}, K., \& {Xilouris}, E.~M.
  2010, \apj, 715, 775

\bibitem[{{Pensabene} {et~al.}(2020){Pensabene}, {Carniani}, {Perna}, {Cresci},
  {Decarli}, {Maiolino}, \& {Marconi}}]{pensabene2020}
{Pensabene}, A., {Carniani}, S., {Perna}, M., {et~al.} 2020, \aap, 637, A84

\bibitem[{{Planck Collaboration} {et~al.}(2020){Planck Collaboration},
  {Aghanim}, {Akrami}, {Ashdown}, {Aumont}, {Baccigalupi}, {Ballardini},
  {Banday}, {Barreiro}, {Bartolo}, {Basak}, {Battye}, {Benabed}, {Bernard},
  {Bersanelli}, {Bielewicz}, {Bock}, {Bond}, {Borrill}, {Bouchet}, {Boulanger},
  {Bucher}, {Burigana}, {Butler}, {Calabrese}, {Cardoso}, {Carron},
  {Challinor}, {Chiang}, {Chluba}, {Colombo}, {Combet}, {Contreras}, {Crill},
  {Cuttaia}, {de Bernardis}, {de Zotti}, {Delabrouille}, {Delouis}, {Di
  Valentino}, {Diego}, {Dor{\'e}}, {Douspis}, {Ducout}, {Dupac}, {Dusini},
  {Efstathiou}, {Elsner}, {En{\ss}lin}, {Eriksen}, {Fantaye}, {Farhang},
  {Fergusson}, {Fernandez-Cobos}, {Finelli}, {Forastieri}, {Frailis},
  {Fraisse}, {Franceschi}, {Frolov}, {Galeotta}, {Galli}, {Ganga},
  {G{\'e}nova-Santos}, {Gerbino}, {Ghosh}, {Gonz{\'a}lez-Nuevo}, {G{\'o}rski},
  {Gratton}, {Gruppuso}, {Gudmundsson}, {Hamann}, {Handley}, {Hansen},
  {Herranz}, {Hildebrandt}, {Hivon}, {Huang}, {Jaffe}, {Jones}, {Karakci},
  {Keih{\"a}nen}, {Keskitalo}, {Kiiveri}, {Kim}, {Kisner}, {Knox},
  {Krachmalnicoff}, {Kunz}, {Kurki-Suonio}, {Lagache}, {Lamarre}, {Lasenby},
  {Lattanzi}, {Lawrence}, {Le Jeune}, {Lemos}, {Lesgourgues}, {Levrier},
  {Lewis}, {Liguori}, {Lilje}, {Lilley}, {Lindholm}, {L{\'o}pez-Caniego},
  {Lubin}, {Ma}, {Mac{\'\i}as-P{\'e}rez}, {Maggio}, {Maino}, {Mandolesi},
  {Mangilli}, {Marcos-Caballero}, {Maris}, {Martin}, {Martinelli},
  {Mart{\'\i}nez-Gonz{\'a}lez}, {Matarrese}, {Mauri}, {McEwen}, {Meinhold},
  {Melchiorri}, {Mennella}, {Migliaccio}, {Millea}, {Mitra},
  {Miville-Desch{\^e}nes}, {Molinari}, {Montier}, {Morgante}, {Moss}, {Natoli},
  {N{\o}rgaard-Nielsen}, {Pagano}, {Paoletti}, {Partridge}, {Patanchon},
  {Peiris}, {Perrotta}, {Pettorino}, {Piacentini}, {Polastri}, {Polenta},
  {Puget}, {Rachen}, {Reinecke}, {Remazeilles}, {Renzi}, {Rocha}, {Rosset},
  {Roudier}, {Rubi{\~n}o-Mart{\'\i}n}, {Ruiz-Granados}, {Salvati}, {Sandri},
  {Savelainen}, {Scott}, {Shellard}, {Sirignano}, {Sirri}, {Spencer},
  {Sunyaev}, {Suur-Uski}, {Tauber}, {Tavagnacco}, {Tenti}, {Toffolatti},
  {Tomasi}, {Trombetti}, {Valenziano}, {Valiviita}, {Van Tent}, {Vibert},
  {Vielva}, {Villa}, {Vittorio}, {Wandelt}, {Wehus}, {White}, {White},
  {Zacchei}, \& {Zonca}}]{planck2018}
{Planck Collaboration}, {Aghanim}, N., {Akrami}, Y., {et~al.} 2020, \aap, 641,
  A6

\bibitem[{{Rizzo} {et~al.}(2022){Rizzo}, {Kohandel}, {Pallottini}, {Zanella},
  {Ferrara}, {Vallini}, \& {Toft}}]{rizzo2022}
{Rizzo}, F., {Kohandel}, M., {Pallottini}, A., {et~al.} 2022, \aap, 667, A5

\bibitem[{{Roman-Oliveira} {et~al.}(2023){Roman-Oliveira}, {Fraternali}, \&
  {Rizzo}}]{roman2023}
{Roman-Oliveira}, F., {Fraternali}, F., \& {Rizzo}, F. 2023, \mnras, 521, 1045

\bibitem[{{Roos} {et~al.}(2015){Roos}, {Juneau}, {Bournaud}, \&
  {Gabor}}]{roos2015}
{Roos}, O., {Juneau}, S., {Bournaud}, F., \& {Gabor}, J.~M. 2015, \apj, 800, 19

\bibitem[{{Rupke} {et~al.}(2005){Rupke}, {Veilleux}, \& {Sanders}}]{rupke2005}
{Rupke}, D.~S., {Veilleux}, S., \& {Sanders}, D.~B. 2005, \apjs, 160, 115

\bibitem[{{Sbarrato} {et~al.}(2021){Sbarrato}, {Ghisellini}, {Giovannini}, \&
  {Giroletti}}]{sbarrato2021}
{Sbarrato}, T., {Ghisellini}, G., {Giovannini}, G., \& {Giroletti}, M. 2021,
  \aap, 655, A95

\bibitem[{{Shao} {et~al.}(2022){Shao}, {Wang}, {Weiss}, {Wagg}, {Carilli},
  {Strauss}, {Walter}, {Cox}, {Fan}, {Menten}, {Narayanan}, {Riechers},
  {Bertoldi}, {Omont}, \& {Jiang}}]{shao2022}
{Shao}, Y., {Wang}, R., {Weiss}, A., {et~al.} 2022, \aap, 668, A121

\bibitem[{{Solomon} \& {Vanden Bout}(2005)}]{solomon2005}
{Solomon}, P.~M. \& {Vanden Bout}, P.~A. 2005, \araa, 43, 677

\bibitem[{{Stanley} {et~al.}(2019){Stanley}, {Jolly}, {K{\"o}nig}, \&
  {Knudsen}}]{stanley2019}
{Stanley}, F., {Jolly}, J.~B., {K{\"o}nig}, S., \& {Knudsen}, K.~K. 2019, \aap,
  631, A78

\bibitem[{{Trinca} {et~al.}(2022){Trinca}, {Schneider}, {Valiante}, {Graziani},
  {Zappacosta}, \& {Shankar}}]{trinca2022}
{Trinca}, A., {Schneider}, R., {Valiante}, R., {et~al.} 2022, \mnras, 511, 616

\bibitem[{{Tripodi} {et~al.}(2023{\natexlab{a}}){Tripodi}, {Feruglio},
  {Kemper}, {Civano}, {Costa}, {Elvis}, {Bischetti}, {Carniani}, {Di Mascia},
  {D'Odorico}, {Fiore}, {Gallerani}, {Ginolfi}, {Maiolino}, {Piconcelli},
  {Valiante}, \& {Zappacosta}}]{tripodi2023b}
{Tripodi}, R., {Feruglio}, C., {Kemper}, F., {et~al.} 2023{\natexlab{a}},
  \apjl, 946, L45

\bibitem[{{Tripodi} {et~al.}(2023{\natexlab{b}}){Tripodi}, {Lelli}, {Feruglio},
  {Fiore}, {Fontanot}, {Bischetti}, \& {Maiolino}}]{tripodi2023a}
{Tripodi}, R., {Lelli}, F., {Feruglio}, C., {et~al.} 2023{\natexlab{b}}, \aap,
  671, A44

\bibitem[{{Tsukui} \& {Iguchi}(2021)}]{tsukui2021}
{Tsukui}, T. \& {Iguchi}, S. 2021, Science, 372, 1201

\bibitem[{{Valentini} {et~al.}(2021){Valentini}, {Gallerani}, \&
  {Ferrara}}]{valentini2021}
{Valentini}, M., {Gallerani}, S., \& {Ferrara}, A. 2021, \mnras, 507, 1

\bibitem[{{Venemans} {et~al.}(2020){Venemans}, {Walter}, {Neeleman}, {Novak},
  {Otter}, {Decarli}, {Ba{\~n}ados}, {Drake}, {Farina}, {Kaasinen},
  {Mazzucchelli}, {Carilli}, {Fan}, {Rix}, \& {Wang}}]{venemans2020}
{Venemans}, B.~P., {Walter}, F., {Neeleman}, M., {et~al.} 2020, \apj, 904, 130

\bibitem[{{Vietri} {et~al.}(2022){Vietri}, {Misawa}, {Piconcelli}, {Franzetti},
  {Luminari}, {Travascio}, {Bischetti}, {Bisogni}, {Bongiorno}, {Bruni},
  {Feruglio}, {Giunta}, {Nicastro}, {Saccheo}, {Testa}, {Tombesi}, {Vignali},
  {Zappacosta}, \& {Fiore}}]{vietri2022}
{Vietri}, G., {Misawa}, T., {Piconcelli}, E., {et~al.} 2022, \aap, 668, A87

\bibitem[{{Volonteri}(2010)}]{volonteri2010}
{Volonteri}, M. 2010, \aapr, 18, 279

\bibitem[{{Volonteri} {et~al.}(2023){Volonteri}, {Habouzit}, \&
  {Colpi}}]{volonteri2023}
{Volonteri}, M., {Habouzit}, M., \& {Colpi}, M. 2023, \mnras, 521, 241

\bibitem[{{Wang} {et~al.}(2019){Wang}, {Wang}, {Fan}, {Wu}, {Yang}, {Neri}, \&
  {Yue}}]{wang2019}
{Wang}, F., {Wang}, R., {Fan}, X., {et~al.} 2019, \apj, 880, 2

\bibitem[{{Wu} {et~al.}(2015){Wu}, {Wang}, {Fan}, {Yi}, {Zuo}, {Bian}, {Jiang},
  {McGreer}, {Wang}, {Yang}, {Yang}, {Thompson}, \& {Beletsky}}]{wu2015}
{Wu}, X.-B., {Wang}, F., {Fan}, X., {et~al.} 2015, \nat, 518, 512

\bibitem[{{Zana} {et~al.}(2022){Zana}, {Gallerani}, {Carniani}, {Vito},
  {Ferrara}, {Lupi}, {Di Mascia}, \& {Barai}}]{zana2022}
{Zana}, T., {Gallerani}, S., {Carniani}, S., {et~al.} 2022, \mnras, 513, 2118

\bibitem[{{Zappacosta} {et~al.}(2023){Zappacosta}, {Piconcelli}, {Fiore},
  {Saccheo}, {Valiante}, {Vignali}, {Vito}, {Volonteri}, {Bischetti},
  {Comastri}, {Done}, {Elvis}, {Giallongo}, {La Franca}, {Lanzuisi},
  {Laurenti}, {Miniutti}, {Bongiorno}, {Brusa}, {Civano}, {Carniani},
  {D'Odorico}, {Feruglio}, {Gallerani}, {Gilli}, {Grazian}, {Guainazzi},
  {Marinucci}, {Menci}, {Middei}, {Nicastro}, {Puccetti}, {Tombesi}, {Tortosa},
  {Testa}, {Vietri}, {Cristiani}, {Haardt}, {Maiolino}, {Schneider}, {Tripodi},
  {Vallini}, \& {Vanzella}}]{zappacosta2023}
{Zappacosta}, L., {Piconcelli}, E., {Fiore}, F., {et~al.} 2023, \aap, 678, A201

\end{thebibliography}
\bibliographystyle{aa}

\end{document}